\documentclass[12pt]{article}
\usepackage{amsmath,amssymb,exscale,graphicx,hyperref}
\usepackage[numbers,sort&compress]{natbib}
\usepackage{color}
\input epsf
\newcommand{\m}[1]{\marginpar{{\tiny *}} }
\def\be{\begin{equation}}
\def\ee{\end{equation}} 

\def\ba#1\ea{\begin{align}#1\end{align}} 
\def\U1FN{U(1)$_{\rm FN}$}
\newcommand{\smalloverleftrightarrow}[1]{%
  \overset{\scriptstyle\leftrightarrow}{#1}%
}

\begin{document}
\topmargin -1.0cm
\oddsidemargin -0.8cm
\evensidemargin -0.8cm

\topmargin -1.0cm
\oddsidemargin -0.8cm
\evensidemargin -0.8cm

\begin{center}
\vspace{40pt}

\Large \textbf{Lepton flavor from %Froggatt-Nielsen mechanism 
a horizontal symmetry\\ in a slice of AdS$_5$}

\end{center}

\vspace{15pt}
\begin{center}
{\bf Leandro Da Rold$^{\star}$, Franco A. Gigena$^{\circ}$, Jaime S. Guzm\'an Guerrero$^{\dagger}$} 

\vspace{20pt}

\textit{Centro At\'omico Bariloche, Instituto Balseiro and CONICET}
\\[0.2cm]
\textit{Av.\ Bustillo 9500, 8400, S.\ C.\ de Bariloche, Argentina}

\end{center}

\vspace{20pt}
\begin{center}
\textbf{Abstract}
\end{center}
\vspace{5pt} {\small \noindent
We build a model of lepton flavor in a slice of AdS$_5$. We add to the 5D SM fields a set of neutrino fields, as well as a horizontal U(1) symmetry and a flavon field, all propagating in the bulk. The electroweak and U(1) symmetries are spontaneously broken by a potential localized on the infrared boundary. We show that in a flavor anarchic scenario, by suitable choice of the 5D masses and the U(1) charges, the masses of the SM leptons and the PMNS matrix can be naturally generated. The neutrino masses are Dirac like, with a normal ordered hierarchical spectrum, $\Delta m_{32}^2\approx m_3^2$, $\Delta m_{21}^2\approx m_2^2\gg m_1^2$, and a suppressed $\theta_{13}$ mixing angle. 
We find configurations where the charged lepton mixing angles are suppressed by powers of the Cabibbo angle ($\lambda_C$) relative to vanilla anarchic partial compositeness, consequently reducing CP and lepton flavor violation. Specifically, the Wilson coefficient for the electron electromagnetic dipole moment exhibits $\lambda_C^2$ suppression, while those governing $\mu\to e\gamma$ and $\mu-e$ vector operators are suppressed by $\lambda_C^{3/2}$ compared to the anarchic scenario without U(1) horizontal symmetry.
%We find configurations for which the mixing angles in interactions of charged leptons are suppressed by powers of $\lambda_C$ compared with anarchic partial compositeness,
%decreasing the amount of CP and lepton flavor violation.
%For these configurations CP and lepton flavor violating processes are supressed compared with the anarchic scenario. 
%The Wilson coefficient of the electron electromagnetic dipole moment is suppressed by $\lambda_C^2$, whereas those of $\mu\to e\gamma$ and $\mu-e$ vector operators are suppressed by $\lambda_C^{3/2}$ compared with the anarchic scenario. 
}

\vfill
\noindent {\footnotesize E-mails:
$\star$ daroldl@ib.edu.ar, 
$\circ$ franco.gigena@ib.edu.ar, 
$\dagger$ jaime.guzman@ib.edu.ar
}

\noindent
\eject

\tableofcontents

\section{Introduction}
The most widely considered solutions to the electroweak (EW) hierarchy problem that rely on symmetries introduce new physics (NP) at the TeV scale~\cite{Witten:1981nf,Martin:1997ns,Kaplan:1983fs,Kaplan:1983sm}, with profound implications for flavor physics.
For generic flavor structure of the NP sector the new states can mediate flavor transition processes with operators suppressed by the TeV scale, in severe contradiction with bounds from the experiments that require the scale of these operators to be of order $10^{2-5}\,{\rm TeV}$~\cite{Isidori:2010kg}. In most scenarios the hierarchy of masses and mixing angles of the Standard Model (SM) fermions is not explained.  
%
%In the abscence of flavor symmetries in the NP sector, the new states can mediate flavor transition processes with operators suppressed by the TeV scale, in severe contradiction with bounds from the experiments that require the scale of these operators to be of order $10^{3-5}\,{\rm TeV}$.

%There are at least two different mechanisms that can explain the hierarchical flavor structure of the SM, as well as suppress the flavor violating operators. First, there is the mechanism of 

There are several approaches to the flavor problem in these theories. In the framework of composite Higgs models, the partial compositeness mechanism, where the SM fermions have linear interactions with the composite sector, can generate the hierarchy of masses and mixings of the charged SM fermions~\cite{Kaplan:1991dc,Grossman:1999ra,Huber:2000ie,Gherghetta:2000qt}. An explanation of neutrino masses and mixing angles in anarchic partial compositeness (APC) usually requires new symmetries~\cite{Csaki:2008qq}. The chiral couplings suppressing the mass and mixing angles also suppress the flavor violating operators, realizing an approximate Glashow-Iliopoulos-Maiani (GIM) mechanism~\cite{Glashow:1970gm}. 
This framework has been thoroughly studied in warped extra dimensions~\cite{Agashe:2004cp}, as well as in two site models~\cite{Contino:2006nn} and in the context of the effective description of the strongly interacting light Higgs~\cite{Giudice:2007fh}.  
It has been shown that, despite the GIM-like suppression, for APC the masses of the resonances of the NP sector, $m_*$, have to be pushed above the TeV. In the quark sector the strongest constraint arises from Kaon physics, with $m_*\gtrsim 20\,{\rm TeV}$~\cite{Csaki:2008zd}. In the lepton sector the most sever constraint is the bound from the electromagnetic dipole moment (EDM) of the electron, that demands $m_*/g_*\gtrsim 200$~TeV, where $g_*$ is the coupling between resonances~\cite{Glioti:2024hye}.

On a different direction, abelian horizontal symmetries can also provide a rationale for the flavor structure of the leptons of the SM~\cite{Davidson:1979wr,Davidson:1983fy,Grossman:1998jj}. 
In the Froggatt-Nielsen (FN) framework, the Standard Model fermions and a new scalar flavon field $\Phi$ are charged under an additional U(1)$_{\rm FN}$ symmetry~\cite{Froggatt:1978nt,Leurer:1992wg,Leurer:1993gy,King:1999mb}. 
Yukawa interactions arise from higher dimensional operators with powers of $\Phi$ that are determined by the charges of the fermions, $p_f$. When $\Phi$ acquires a vacuum expectation value, spontaneously breaking U(1)$_{\rm FN}$, a flavor structure is generated with the effective Yukawa couplings being suppressed by $(\langle\Phi\rangle/\Lambda)^{|p_L-p_R|}$. FN provides a small parameter and an organizing principle for flavor, even with anarchic Yukawa couplings in the fundamental theory~\cite{Haba:2000be}. FN scenarios for the lepton flavor~\cite{Suematsu:2001sp,Dreiner:2003hw,Nir:2004my,Kamikado:2008jx,Plentinger:2008up,Buchmuller:2011tm,Krippendorf:2015kta,Bauer:2016rxs,Ema:2016ops,Nishiwaki:2016xyp,Feruglio:2019ybq,Berger:2019rfk,Smolkovic:2019jow,Nishimura:2020nre,Qiu:2023igq} have been studied in supersymmetric theories~\cite{Sato:2000ff}, in the presence of leptoquarks~\cite{Bordone:2019uzc,Asadi:2023ucx}, in the framework of modular symmetry~\cite{Pathak:2025fpo} and in the context of elementary Axiflavon-Higgs unification~\cite{Alanne:2018fns}. Recently refs.~\cite{Ibe:2024cvi,Cornella:2024jaw} identified viable charge assignments and studied lepton flavor violation (LFV). Several challenges remain, as the quest for the potential of the scalar fields, the need of many new heavy fermions and potential issues given the presence of an axion~\cite{Calibbi:2016hwq}. 

%The present paper focuses on the flavor of the leptonic sector. One of the goals is the development of a natural model capable of simultaneously explaining the hierarchy of the charged lepton masses, the neutrino mass-squared difference hierarchy, and the PMNS matrix featuring two large mixing angles and a small $\theta_{13}$. 

This work addresses the flavor structure of the leptonic sector in APC with a horizontal U(1) symmetry~\footnote{See Ref.~\cite{DaRold:2021cca} for a model of quark flavor incorporating both mechanisms.}. We pursue two primary objectives. First, we aim to construct a natural model that simultaneously explains the charged lepton mass hierarchy, the neutrino mass-squared difference hierarchy, and the PMNS matrix, which includes two large mixing angles and a small one. It is highly challenging for APC or models with horizontal symmetries to individually account for these hierarchies. Second we seek to suppress LFV processes, thereby allowing the new physics scale to be lowered compared with vanilla APC (hereafter simply APC).

We construct a flavor-anarchic 5D model in a slice of AdS$_5$, with the Standard Model fields propagating in the bulk. This setup is extended by a 5D U(1)$_{\rm FN}$ gauge symmetry, a 5D flavon field and three 5D neutrino fields that are SM singlets. All the the fermions are charged under the new symmetry. Electroweak and U(1)$_{\rm FN}$ breaking is achieved via an IR boundary potential, generating masses for the lepton 0-modes. The light neutrinos are of Dirac nature and, by appropriate choice of FN charges and 5D bulk masses, the observed hierarchies of masses and mixing angles are naturally reproduced, with $U_{\rm PMNS}\simeq U_{\nu_L}$.

%We will consider a theory of lepton flavor that combines both mechanisms~\footnote{See Ref.~\cite{2112.14600} for a model of flavor for the quark sector with both mechanisms.}. We will consider a flavor anarchic 5D theory in a slice of AdS$_5$, with the SM fields propagating in the extra dimension. We will add a 5D U(1)$_{\rm FN}$ gauge symmetry, as well as a 5D flavon scalar charged under U(1)$_{\rm FN}$. We will show that a potential on the IR boundary can break the EW and U(1)$_{\rm FN}$ symmetries, generating masses for the 0-mode leptons. We will consider a scenario where the light neutrinos have Dirac-type masses, and we will show that by properly choosing the FN charges and the 5D masses of the lepton fields one can obtain the lepton masses and the PMNS matrix, with $U_{\rm PMNS}\simeq U_{\nu_L}$. 

We study CP and lepton flavor violation (LFV) and compare the predictions with APC~\cite{Agashe:2006iy,Beneke:2015lba}. As in that case, the largest contributions to these processes arise from vector and dipole operators, respectively generated at tree and one-loop level. 
%In APC the leading contributions to chirality flipping operators are of $Y^3$, since operators linear in the Yukawa are approximately aligned with the masses~\cite{Agashe:2004cp}. However in models with FN symmetry there are new sources of flavor violation, namely the U(1)$_{\rm FN}$ charges: $P_{L,E,N}$, that are generation dependent. Therefore one has to study chirality flipping operators linear in the Yukawa coupling, as those of ${\cal O}(P^2Y)$.
We find that CP and LFV dipole operators are dominated by one-loop contributions with a heavy spin one resonance and with either a neutral or a charged Higgs. 
%that are of order $y_*^3$, and, that are of order $g_*^2y_*$, where $y_*$ and $g_*$ are the couplings with the Higgs field and with the spin one resonances, respectively. 
In the first case contributions are enhanced by the FN-charge squared, disfavoring large charges. We show that for suitable choices of 5D masses and charges the Wilson coefficient of dipole and vector operators are suppressed by powers of $\lambda_C$, compared with APC.% Although this suppression relaxes the flavor bounds, the eEDM still requires $m_*/g_*\sim 20-40\,{\rm TeV}$.

The paper is organized as follows: in sec.~\ref{sec-model} we define a 5D model of leptonic flavor and determine a texture for the neutrino mass matrix. In sec.~\ref{sec-symb} we describe the scalar sector of the model, showing the conditions for spontaneous symmetry breaking and performing the Kaluza-Klein (KK) decomposition of the neutral fields. In sec.~\ref{sec-flavor} we study the leptonic sector of the model, we perform the KK decomposition showing how to obtain the mass matrices of the light leptons and their interactions with the resonances. In sec.~\ref{sec-LFV} we study LFV and in sec.~\ref{sec-numbers} we show the numerical predictions. We present some discussions and conclusions in sec.~\ref{sec-conclusions}. Some details regarding the KK decomposition of the scalar fields and the diagonalization of the fermion mass matrices are provided in Appendices~\ref{app-scalar} and~\ref{ap-diag}.

\section{A 5D model of leptons with horizontal U(1) symmetry}
\label{sec-model}
We consider a slice of AdS$_5$ and work in conformal coordinates, with the interval $ds^2=a(z)^2(\eta_{\mu\nu}dx^\mu dx^\nu-dz^2)$, and the warp factor $a(z)=L/z$. The extra dimension is compactified on an interval with UV and IR boundaries at $z_0\sim 10^{-15}$~TeV$^{-1}$ and $z_1\sim$~TeV$^{-1}$, respectively~\cite{Randall:1999ee}.

The Lagrangian of the 5D theory describing the flavor of the leptons is:
\be
S_5= \frac{1}{g_5^2} \int d^4x \int_{z_0}^{z_1} dz ({\cal L}_{\rm g} + {\cal L}_{f} + {\cal L}_{s}+{\cal L}_{y}) 
\ee
where the different terms correspond respectively to gauge, fermion and scalar sectors, as well as the Yukawa interactions. 
$\cal L_{\rm g}$ contains the kinetic term of the 5D gauge symmetry: SU(2)$_L\times$U(1)$_Y\times$U(1)$_{\rm FN}$.
%$\cal L_{\rm g}$ contains the 5D kinetic term: $F_{MN}F^{MN}$, of the SM gauge group SU(2)$_L\times$U(1)$_Y$, as well as the U(1)$_{\rm FN}$ field. 
The boundary conditions (BCs) of the SM and FN gauge fields are chosen as $(++)$ and $(-+)$, respectively, leading to 0-modes that correspond to the 4D SM gauge fields only~\footnote{$(+)$ and $(-)$ denote Neumann and Dirichlet BCs, respectively, the first sign is for the UV BC and the second one for the IR BC.}. We have factorized a coefficient $1/g_5^2$, with $g_5$ being the 5D expansion parameter~\footnote{While we use a single common coupling $g_5$ for brevity, distinct couplings may be assigned to each Lagrangian term. This is required, for example, to match the 4D gauge couplings.}.

${\cal L}_{f}$ contains the kinetic and mass terms of the 5D fermions, with one field for each chiral lepton of the SM, plus a 5D neutrino field.
%\be
%{\cal L}_{f} = \frac{\sqrt{g}}{g_5^2} \sum_{f=L,E,N}\bar f(i\Dslash-m_f)f \ .
%\ee
We will denote the SU(2)$_L$ doublets by ${\rm L}=({\rm L}^N,{\rm L}^E)$ and the singlets by $E$ and $N$. The 5D masses will be written as: $m_\Psi=c_\Psi/L$, with $\Psi={\rm L},E,N$ and $c_\Psi\sim{\cal O}(1)$. The BCs ensure that every 5D field yields a 0-mode for each SM chiral fermion and a right-handed neutrino $N_R^{(0)}$ per generation. The charges of the leptons under the SM gauge group are the usual ones, whereas their charges under U(1)$_{\rm FN}$ are denoted as $p_{{\rm L},E,N}$, see table~\ref{t-rep}. The FN charges of the leptons are not universal, we will elaborate on them on sec.~\ref{sec-flavor}

$\cal L_{\rm s}$ includes the 5D scalars $H$ and $\Phi$, with 0-modes identified as the 4D scalar Higgs and flavon fields, respectively:
\be
{\cal L}_s = \sqrt{g} \left(|D_MH|^2-m_H^2|H|^2+|D_M\Phi|^2-m_\Phi^2|\Phi|^2\right)\ ,%x_\nu \bar L \tilde H N \left(\frac{\Phi^{(\dagger)}}{\Lambda_5}\right)^{\beta_N}+x_e \bar L H E\left(\frac{\Phi^{(\dagger)}}{\Lambda_5}\right)^{\beta_E}\right]
\ee
where $g$ is the determinant of the 5D metric and $M$ labels the components in 5D: $(\mu,5)$. We assume that the Higgs field is not charged under U(1)$_{\rm FN}$, whereas $\Phi$ has charge $p_\Phi=1/2$, as summarized in table~\ref{t-rep}. The BCs are taken as $(++)$ for both fields.

${\cal L}_y$ is given by:
\ba
& {\cal L}_y = \sqrt{g} \left[x_N \bar {\rm L} \tilde H N \left(\frac{\Phi^{(\dagger)}}{\Lambda_5}\right)^{\alpha_N}+x_E \bar {\rm L} H E\left(\frac{\Phi^{(\dagger)}}{\Lambda_5}\right)^{\alpha_E}\right] + {\rm h.c.} \ ,
\label{eq-Ly}
\\
& \alpha_N=2|p_{\rm L}-p_N|\ , \qquad  \alpha_E=2|p_{\rm L}-p_E| \ ,
\ea
where generation indices are understood for the fields as well as for the couplings and the exponents $\alpha_{N,E}$. 
The choice between $\Phi$ and $\Phi^\dagger$ in the interactions is determined by the sign of $(p_{\rm L} - p_N)$ and $(p_{\rm L} - p_E)$, ensuring that the Yukawa couplings remain invariant under the U(1)$_{\rm FN}$ symmetry. $\Lambda_5$ denotes the cut-off scale of the 5D theory, with $\Lambda_5 \sim 24\pi^3 / g_5^2$. The couplings $x_{N,E}$ are dimensionless $\mathcal{O}(1)$ parameters, assumed to be anarchic in flavor space. Upon canonical normalization of the fields, the 5D Yukawa couplings take the form $x_{N,E} g_5$. As is typical in FN models, replacing $\Phi$ with its vacuum expectation value (vev) induces effective 5D Yukawa couplings with a non-trivial flavor structure, governed by the FN charges: $x_{N,E} (v_\Phi/\Lambda_5)^{2|p_{\rm L}-p_{N,E}|}$.

\begin{table}[ht]
\centering
\begin{tabular}{|c|c|c|c|}
\hline\rule{0mm}{5mm}
field & SU(2)$_L$ & U(1)$_Y$ & U(1)$_{\rm FN}$ \\[5pt]
\hline 
$H$ & \bf2 & $1/2$ & 0\\[3pt]
\hline 
$\Phi$ & \bf1 & 0 & $1/2$ \\[3pt]
\hline 
${\rm L}$ & \bf2 & $1/2$ & $p_L$\\[3pt]
\hline 
$E$ & \bf1 & $-1$ & $p_E$\\[3pt]
\hline 
$N$ & \bf1 & 0 & $p_N$\\[3pt]
\hline 
\end{tabular}
\caption{Representations of the the 5D leptons and scalar fields.}
\label{t-rep}
\end{table}

The UV and IR boundaries host quadratic and quartic potentials in the scalar fields, respectively:
\ba
%& S_1= \int d^4x \ {\cal L}_1
%\\
%& {\cal L}_1 = -[\sqrt{g_4} (m_{1H}^2|H|^2+m_{1\Phi}^2|\Phi|^2+\lambda_{1H}|H|^4+\lambda_{1\Phi}|\Phi|^4+\lambda_{1\Phi H}|\Phi|^2|H|^2)]|_1
& S_{\rm bd}= \int d^4x \ \left[a^4(z_0)\ V_0 - a^4(z_1)\ V_1\right] \ ,
\\
& V_0 = (m_{0H}^2|H|^2+m_{0\Phi}^2|\Phi|^2)|_{z_0} \ ,
\\
& V_1 = (m_{1H}^2|H|^2+m_{1\Phi}^2|\Phi|^2+\lambda_{1H}|H|^4+\lambda_{1\Phi}|\Phi|^4+\lambda_{1\Phi H}|\Phi|^2|H|^2)|_{z_1} \ ,
\ea
the subscripts 0 and 1 refer to the UV and IR boundaries at $z_0$ and $z_1$, respectively.

\subsection{Texture of the neutrino mass matrix}
We will consider Dirac-type neutrinos, with normal ordering and a PMNS matrix dominated by the neutrino sector, meaning that the mixing angles of the Left-handed charged leptons are much smaller than those of the neutral ones. This feature enables the suppression LFV processes.
The masses and mixing angles of the SM neutrinos are determined with high accuracy by the mass matrix of the would be 0-modes obtained after dimensional reduction, neglecting the effect of mixing with the KK states.
%Let us focus now on the scenario of neutrino masses and mixing angles in the 4D theory, after dimensional reduction, keeping only the would be 0-modes. 

We find it useful to express the hierarchy of masses and mixing angles in powers of $\lambda_C$.
Although the power of $\lambda_C$ featuring angles and ratios of masses is not an exact number, and it depends on the renormalization scale, we will consider approximate values valid at a few TeV scale~\cite{Antusch:2013jca}.
For the hierarchical normal order of neutrino masses we will use:
\be
m_{\nu 1}\sim\lambda_C^2\ m_{\nu 3} \ , \qquad\qquad m_{\nu 2}\sim\lambda_C\ m_{\nu 3} \ .
\label{eq-mnu}
\ee
The PMNS matrix has a non-hierarchical structure, except for $\theta_{13}\simeq 0.15\sim\lambda_C$, whereas $\theta_{12}\sim\theta_{23}\sim{\cal O}(1)$. Since we are interested in the limit: $U_{\rm PMNS}\simeq U_{\nu_L}$, we will work with~\cite{Hirsch:2001he}
\be
U_{\nu_L}\sim \left(\begin{array}{ccc} 1 & 1 & \lambda_C \\ 1 & 1 & 1 \\ 1 & 1 & 1\end{array}\right) \ .
\label{eq-unu}
\ee

The following neutrino mass matrix satisfies the previous conditions:
\be
M_\nu\sim m_{\nu 3}\left(\begin{array}{ccc} \lambda_C^2 & \lambda_C & \lambda_C \\ \lambda_C^2 & \lambda_C & 1 \\ \lambda_C^2 & \lambda_C & 1\end{array}\right) \ .
\label{eq-mnu1}
\ee

For the hierarchy of charged lepton masses we consider the following expressions:
\be
\label{eq-emass}
%m_e\sim g_* v \lambda_C^9 \ , \qquad m_\mu \sim g_* v \lambda_C^6 \ , \qquad m_\tau\sim g_* v \lambda_C^4 \ ,
m_e\sim \lambda_C^5 m_\tau \ , \qquad\qquad m_\mu \sim \lambda_C^2 m_\tau \ .
\ee

In sec.~\ref{sec-flavor} we will show configurations of 5D masses and FN charges that in the 0-mode approximation lead to Eqs.~(\ref{eq-mnu1}) and~(\ref{eq-emass}), as well as small mixing angles in the charged sector.

\section{Scalar sector of the 5D model and symmetry breaking}
\label{sec-symb}
Let us now consider the scalar sector of the model. We focus on the radial mode of the scalars, describing the vevs and the first KK state~\footnote{The angular mode of $\Phi$ can lead to a 4D axion, that can obtain a mass from the KK gluons~\cite{Gherghetta:2020keg}, the analysis of the angular modes is beyond the scope of this paper.}.
We present the results that are relevant for the theory of leptonic flavor. The details of the calculations of the potential, as well as of the KK decomposition, are given in Ap.~\ref{app-scalar}.

The potential localized on the IR boundary, $V_1$, triggers the spontaneous breaking of the EW and FN symmetries, preserving the electromagnetic symmetry. The vevs of the scalar fields, $\langle H^t\rangle=(0,v_H(z))$ and $\langle \Phi\rangle=v_\Phi(z)$, can be written as~\cite{Cacciapaglia:2006mz}:
\be
v_\chi(z)=v_{\chi4}\frac{g_5}{\sqrt{L}}\frac{z_1}{L}\left(\frac{z}{z_1}\right)^{2+\beta_\chi} F(2+2\beta_\chi)\ , \qquad \chi=H,\Phi \ ,
\ee
where $\beta_\chi=\sqrt{4+m_\chi^2L^2}$ and the function $F(x)$ is defined by:
\be
F(x)=\left[\frac{x}{1-(z_0/z_1)^x}\right]^{1/2} \ .
\label{eq-F}
\ee
Notice that this function, given the large hierarchy between $z_0$ and $z_1$, can have an exponential dependence on its argument, with $F(x)\to \sqrt{x}(z_0/z_1)^{x/2}$ for $x<0$, $F(x)\to \sqrt{x}$ for $x>0$ and $F(x)\to [\log(z_1/z_0)]^{-1/2}$ for $x\to 0$.~\footnote{$F$ can be related with the degree of compositeness of two-site theories.}

The 4D vevs are obtained as the integrals of the 5D ones along the extra-dimension:
\be
v_{\chi 4}^2=\frac{1}{g_5^2}\int dz \ a(z)^3 v_\chi(z)^2 \ ,  \qquad \chi=H,\Phi.
\ee

Since the scalar fields are mixed by $\lambda_{1\Phi H}$ in $V_1$, we perform the following KK decomposition:
\be
h(x^\mu,z)=\sum_n s^{(n)}(x^\mu)f^h_n(z) \ , \qquad \phi(x^\mu,z)=\sum_n s^{(n)}(x^\mu)f^\phi_n(z) \ , 
\label{eq-KKs1}
\ee
where $s^{n}(x^\mu)$ are the 4D KK scalar modes, with projections on the 5D fields $h(x^\mu,z)$ and $\phi(x^\mu,z)$ given by the KK profiles $f^h_n(z)$ and $f^{\phi}_n(z)$, respectively. 

\begin{figure}[t]
\centering
\includegraphics[width=0.49\textwidth]{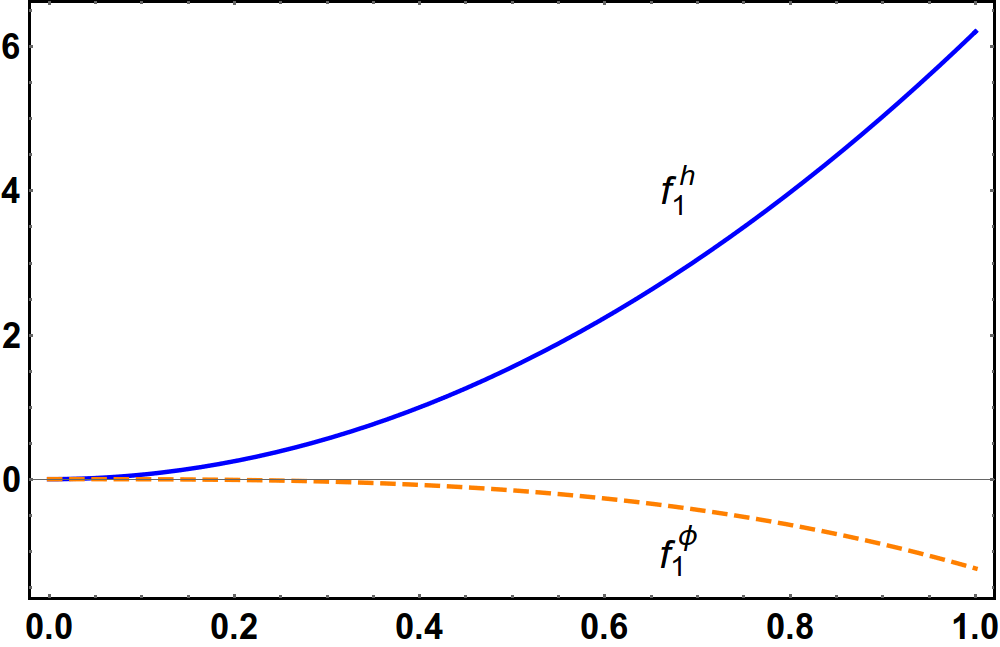}
\includegraphics[width=0.49\textwidth]{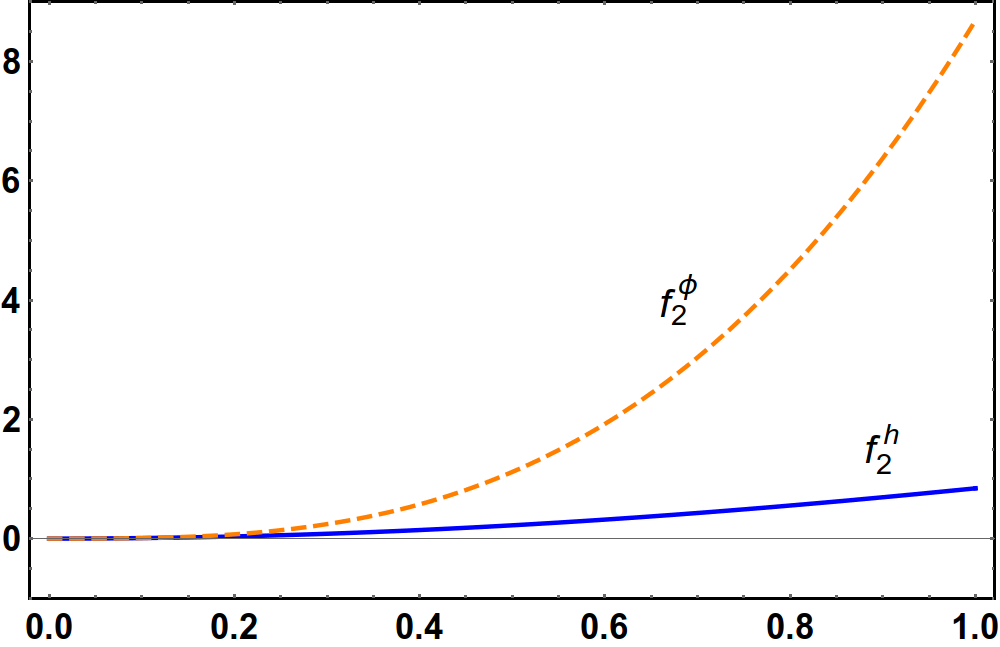}
\caption{KK wave functions of the first and second scalar modes: $s^{(1)}$ and  $s^{(2)}$, respectively.}
\label{fig-KKs}
\end{figure}

We are interested in a region of the parameter space able to generate a light state $s^{(1)}$ that could be identified with the Higgs, with states $s^{(n>1)}$ heavy enough to evade the bounds from searches of scalar resonances. Besides, since the Higgs couplings have been measured in agreement with the SM, the KK wave function of the light state must be dominated by the projection $f^h_1$, with only a small projection $f^\phi_1$. A small coupling $\lambda_{1\Phi H}$ can avoid large mixing between the scalars. 
In Fig.~\ref{fig-KKs} we show the KK wave functions of the first and second modes for a point of the parameter space with KK masses $m_1\simeq 125$~GeV and $m_2\simeq 544$~GeV. The point of the parameter space of the 5D theory is given by: $\beta_H=0$, $\beta_\Phi=1$, $\lambda_{1H}=\lambda_{1\Phi}=\lambda_{1\Phi H}=0.1$, $g_5=\pi\sqrt{2L}$, $\delta m_{1H}^2\simeq -(0.024/L)^2$, $\delta m_{1\Phi}^2\simeq -(0.032/L)^2$ and $z_1=1$~TeV$^{-1}$. 
We find that $f^h_1$ is very well approximated by the 0-mode $f^h_0$ obtained with vanishing quartic coupling and suitable boundary masses. Besides in Ap.~\ref{app-scalar} we provide approximations for a simple calculation of $m_1$ and $m_2$.
A detailed calculation of Higgs couplings and the associated observables must be done for a more quantitative statement about this point of parameter space.

The Higgs interacts with the KK vectors arising from the 5D EW and FN symmetries. For the analysis of flavor violation one needs the coupling with $V^{(0)}$ and $V^{(1)}$:
\be
g^{hh}_{(0n)}=\frac{1}{g_5^2}\int dz\ a(z)^3 f^{V}_0(z)f^{V}_n(z)(f^h_0(z))^2\simeq \frac{g_5}{\sqrt{L}} I^{hh}_{(0n)}(\beta_H)
\ee
where $f^{V}_0(z)$ and $f^{V}_1(z)$ are the KK wave functions of the 0- and 1-modes of a 5D vector, respectively, and $I^{hh}_{(0n)}(\beta_H)\sim{\cal O}(1)$ for $\beta_H\gtrsim -0.5$~\cite{Agashe:2006iy}.

\section{Spectrum and interactions of leptons}
\label{sec-flavor}
Let us focus now on the fermions, as well as their interactions with the Higgs and the vector resonances. We will show that with a suitable choice of 5D parameters one can naturally obtain the masses of the light neutrinos and charged leptons of Eqs.~(\ref{eq-mnu1}) and~(\ref{eq-emass}). We will also identify configurations with small mixing angles in the charged sector, and we will determine the flavor structure of the couplings. 

\subsection{KK decomposition}
We neglect the Yukawa interactions of the 5D fermions and perform a KK decomposition as: $\psi_{L,R}(x^\mu,z)=\sum_n\psi^{(n)}_{L,R}(x^\mu)f^{\psi_{L,R}}_n(z)$. The sum includes a 0-mode only for $(++)$ BCs, leading to a chiral spectrum of 0-modes that reproduce the lepton sector of the SM, plus three Right handed neutrinos: $N^{(0)}_R$. A generation index for each 5D and 4D fermion field is understood.

For the neutral leptons we will consider only the 0-modes, whereas for the charged leptons we will consider the 0-mode and the first resonance of each 5D fermion, grouping them in the following chiral vectors:
\be
\zeta_L^j=({\rm L}^{E(0),j}_L,{\rm L}^{E(1),j}_L,E^{(1),j}_L) \ , \qquad \qquad 
\zeta_R^j=(E^{(0),j}_R,{\rm L}^{E(1),j}_R,E^{(1),j}_R) \ .
\label{eq-zeta}
\ee
Notice that, taking into account generation indices, the vectors $\zeta_{L,R}$ have nine components, we will use a Latin superscript for generations and a Greek subscript for the KK components explicitly shown in Eq.~(\ref{eq-zeta}): $\zeta^j_{L,R\alpha}$.

%The masses and PMNS matrix of the light leptons are generated by the Yukawa interactions of Eq.~(\ref{eq-Ly}). For the calculation of the flavor observables we are also interested in the interactions with the Higgs, as well as with KK vector fields, involving KK fermions. In this section we will show the corresponding couplings.

The 4D Yukawa couplings are obtained after integration over the extra dimension of Eq.~(\ref{eq-Ly}) with the appropriate KK states and the vev of $\Phi$. For the neutral leptons we are interested in the Yukawa coupling involving the 0-modes only:
\be
y_{N}^{jk}= \frac{x_{N}^{jk}}{g_5^2}\int dz\ a(z)^5 f^{{\rm L}_L,j}_{0}(z)f^{N_R,k}_{0}(z) f^h_0(z)\left(\frac{v_{\Phi4}f^\phi_0(z)}{\Lambda_5}\right)^{\alpha_N^{jk}} \ ,
\label{eq-yn}
\ee
whereas for the charged leptons we consider also the first KK state:
\be
y^{jk}_{E}=\left(\begin{array}{ccc} y_{11}^{jk} & 0 & y_{13}^{jk} \\ y_{21}^{jk} & 0 & y_{23}^{jk} \\ 0 & y_{32}^{jk} & 0 \end{array}\right) \ ,
\label{eq-ye}
\ee
with
\be
y^{jk}_{\alpha\beta}= \frac{x_E^{jk}}{g_5^2}\int dz\ a(z)^5 f^{\zeta_L,j}_\alpha(z)f^{\zeta_R,k}_\beta(z) f^h_0(z)\left(\frac{v_{\Phi4}f^\phi_0(z)}{\Lambda_5}\right)^{\alpha_E^{jk}} \ ,
\label{eq-ye1}
\ee
where the KK wave-functions $f^{\psi,j}_\alpha$ are those corresponding to the modes of Eq.~(\ref{eq-zeta}). Notice that, to shorten notation, we omit a subindex $E$ in the components of the charged Yukawa matrix.

The couplings of the charged leptons with the first KK neutral vectors, as $Z^{(1)}$ and $F^{(1)}$, are given by:
\be
g^{jk}_{L}=\delta^{jk}\left(\begin{array}{ccc} g_{L,11}^j & g_{L,12}^j & 0 \\ g_{L,12}^{j} & g_{L,22}^{j} & 0 \\ 0 & 0 & g_{L,33}^{j} \end{array}\right) \ ,
\qquad
g^{jk}_{R}=\delta^{jk}\left(\begin{array}{ccc} g_{R,11}^j & 0 & g_{R,13}^j \\ 0 & g_{R,22}^{j} & 0 \\ g_{R,13}^{j} & 0 & g_{R,33}^{j} \end{array}\right) \ ,
\label{eq-ge}
\ee
where we have factorized from the coupling the non-universal charge: $Q^Z=(T^3-s_{\rm W}^2 Q)$ and $p$, respectively for $Z^{(1)}$ and $F^{(1)}$, and with the following definition:
% over the extra dimension:
\be
g^j_{L,\alpha\beta}= \frac{1}{g_5^2} \int dz\ a(z)^4 f_{\alpha}^{\zeta_{L},j}(z)f_{\beta}^{\zeta_{L},j}(z)f_1^V(z) 
\ , \qquad g^j_{R,\alpha\beta}=g^j_{L,\alpha\beta}(L\to R)
\ .
\label{eq-ge1}
\ee

The flavor structure of the couplings of Eq.~(\ref{eq-ye}) can be simplified by considering the following approximations:
\begin{alignat}{2}
& y^{jk}_{11}= x_E^{jk}\frac{g_5}{\sqrt{L}} F(1-2c_{\rm L}^j)\delta^{\alpha_E^{jk}}F(1+2c_E^k)I_{11}^{jk} \ , 
&\quad & I_{11}^{jk}=\frac{F(2+2\beta_H)F(2+2\beta_\Phi)^{\alpha_E^{jk}}}{F(2-c_{\rm L}^j+c_E^k+\beta_H+\alpha_E^{jk}(2+\beta_\Phi))} \ ,
\nonumber \\
& y^{jk}_{13}\simeq x_E^{jk}\frac{g_5}{\sqrt{L}} F(1-2c_{\rm L}^j)\delta^{\alpha_E^{jk}}I_{13} \ , 
&\quad & I_{13}\approx 0.77 \ ,
\nonumber \\
& y^{jk}_{21}\simeq x_E^{jk}\frac{g_5}{\sqrt{L}} \delta^{\alpha_E^{jk}}F(1+2c_E^k)I_{21} \ , 
&\quad & I_{21}\approx 0.77 \ ,
\nonumber \\
& y^{jk}_{23}\simeq x_E^{jk}\frac{g_5}{\sqrt{L}} \delta^{\alpha_E^{jk}}I_{23} \ , 
&\quad & I_{23}\approx 1 \ ,
\nonumber \\
& y^{jk\dagger}_{32}\simeq x_E^{jk}\frac{g_5}{\sqrt{L}} \delta^{\alpha_E^{jk}}I_{32}^\dagger(\alpha_E^{jk}) \ , 
&\quad & I_{32}\approx \frac{0.6}{1+\alpha_E^{jk}} \ ,
\label{eq-yejk}
\end{alignat}
with:
\be
\delta = \frac{g_5}{\sqrt{L}}\frac{v_{\Phi 4}z_1}{L\Lambda_5} \ .
\label{eq-delta}
\ee
We have checked that corrections to these approximations are below $20\%$ for the region of the parameter space of interest, which will be specified in the next subsection. 
A few comments are in order.
Notice that the 4D Yukawa coupling of Eq.~(\ref{eq-yn}) gives the same result as $y_{11}$ in Eq.~(\ref{eq-yejk}), replacing $E\to N$.
Since the scalars and the KK fermions are localized towards the IR boundary, the integrals with one 0-mode are suppressed when $f_0^\psi$ is localized towards the UV boundary, whereas integrals with two KK fermions $(++)$ are ${\cal O}(1)$ and integrals with two KK modes $(--)$ are suppressed for large $\alpha$, since in this case the support of $(f_0^\phi)^\alpha$ shifts towards the IR boundary where $f_1^{\psi(--)}$ vanishes. 
Note that for a bulk Higgs, the so-called "wrong" Yukawa coupling $y^{jk}_{32}$ does not vanish; instead, it is misaligned from $y^{jk}_{23}$ while sharing the same hierarchical structure.

Let us focus on $y^{jk}_{11}$. The factors $F(2+2\beta_{H,\Phi})$, as well as the denominator of $I_{11}^{jk}$, are of ${\cal O}(1)$ for the region of parameter space of interest. The factors $F(1\mp 2c_{{\rm L},\psi})$ give an exponential suppression if $\pm c_{{\rm L},\psi}>1/2$, as is well known in theories of flavor in a slice of AdS$_5$. The new ingredient is the factor $\delta^{\alpha^{jk}_E}$, that plays the same role as in FN in 4D. By using the estimate for $\Lambda_5$ (see sec.~\ref{sec-model}) and $g_5^2/L=(4\pi)^2/N_{\rm CFT}$, we obtain:
\be
\delta \sim v_{\Phi 4}z_1\frac{8}{3N_{\rm CFT}^{3/2}} \ .
\ee
We will consider 
\be
\delta\approx \sqrt{\lambda_C}\approx 0.47 \ , 
\label{eq-deltan}
\ee
which, for a natural vev: $v_{\Phi 4}z_1\sim 1$, requires a small value $N_{\rm CFT}\simeq 3$, leading to a sizable 5D coupling in the scalar sector.%: $g_5/\sqrt{L}\simeq 7$.

To simplify the flavor structure of the couplings with $V^{(1)}$, (\ref{eq-ge}), it is useful to consider the following approximations:
\begin{alignat}{2}
& g_{L,11}^j \approx {\cal C}_{1,2}\frac{g_5}{\sqrt{L}}%\left(\log\frac{z_1}{z_0}\right)^{1/2}
\frac{5- 2c_{\rm L}^j}{4(2- c_{\rm L}^j)(3- c_{\rm L}^j)}F(1- 2c_{\rm L}^j)^2 \ ,%\left[F(1\mp c_{\psi i})-\frac{1}{\log{z_1/z_0}}\right] \ .
&\quad & g_{R,11}^j = g_{L,11}^j(c_{\rm L}\to -c_E) \ , 
\nonumber \\
& g_{L,12}^j \approx \frac{g_5}{\sqrt{L}}F(1- 2c_{\rm L}^j) \ ,
&\quad & g_{R,13}^j = g_{L,12}^j(c_{\rm L}^j\to -c_E^j) \ ,
\nonumber \\
& g_{L,22}^j \approx \frac{g_5}{\sqrt{L}} \ ,
&\quad & g_{R,22}^j \approx \frac{g_5}{\sqrt{L}} \ 0.77 \ , 
\nonumber \\
& g_{L,33}^j \approx \frac{g_5}{\sqrt{L}} \ 0.77 \ ,
&\quad & g_{R,33}^j \approx \frac{g_5}{\sqrt{L}} \ ,
\label{eq-gni}
\end{alignat}
where, for $g_{L,R,11}$, we omitted a universal term present in vector fields with $(+)$ BC in the UV, and ${\cal C}_{n,2}$ denotes the coefficient of the quadratic term in the expansion around the UV boundary of $f_n^V(z)$: $f_n^V(z)={\cal C}_{n,0}+{\cal C}_{n,2}(z-z_0)^2+\dots$, for which, in the case of the first KK vector, ${\cal C}_{1,2}\approx 3.4$. Corrections to these approximations are below $10\%$ for the region of the parameter space of interest.

\subsection{Light leptons}
\label{sec-lightleptons}
To leading order the masses and mixing angles of the light leptons are obtained from the $3\times 3$ mass matrices:
\be
m_N^{jk}=v_{H4} \ y_{N}^{jk} \ , \qquad \qquad m_E^{jk}=v_{H4} \ y_{11}^{jk} \ ,
\label{eq-m0modes}
\ee
with the 4D Yukawa couplings defined in Eqs.~(\ref{eq-yn}) and~(\ref{eq-yejk}). See Ap.~\ref{ap-diag} for the diagonalization of the mass matrix in KK space.

It is useful to parameterize $m_{N,E}$ in terms of powers of $\lambda_C$, to this end we express the degree of compositeness of the 0-modes in powers of $\lambda_C$, and we define the corresponding exponent through the following relation:
\be
F(1\mp 2c_{{\rm L},\psi})\equiv\lambda_C^{n_{{\rm L},\psi}} \ .
\label{eq-Fn}
\ee
Using Eqs.~(\ref{eq-deltan}) and~(\ref{eq-Fn}) in~(\ref{eq-yn}) and~(\ref{eq-yejk}), one gets:
\be
m_\psi^{jk} \sim v_{H4}\,x_{\psi}^{jk} \frac{g_{5}}{\sqrt{L}}\lambda^{n_{\rm L}^j+|p_{\rm L}^j-p_\psi^k|+n_\psi^k} \ , \qquad \psi=N,E \ .
\label{eq-mpsil}
\ee

The eigenvalues and eigenvectors of the the mass matrices of Eq.~(\ref{eq-mpsil}) can be computed perturbatively in powers of $\lambda_C$. By choosing suitable values of $n_\psi^j$ and $p_{\psi^j}$ the components of the Yukawa matrix have integer powers of $\lambda_C$ and it is straightforward to perform these expansions. Very useful approximations can be obtained by considering the following estimates~\cite{DaRold:2021cca}:
\ba
& m_{\psi j}\sim v_{H4}\,\frac{g_{5}}{\sqrt{L}}\lambda^{n_{\rm L}^j+|p_{\rm L}^j-p_{\psi}^j|+n_{\psi}^j} \ ,
\\ 
& \theta_{\psi_L}^{jk}\sim \lambda^{n_{\rm L}^j-n_{\rm L}^k+|p_{\rm L}^j-p_{\psi}^k|-|p_{\rm L}^k-p_{\psi}^k|} \ , \ j<k \ ,
\label{eq-thetaL}
\\
& \theta_{\psi_R}^{jk}\sim \lambda^{n_\psi^j-n_\psi^k+|p_{\rm L}^k-p_{\psi}^j|-|p_{\rm L}^k-p_{\psi}^k|} \ , \ j<k \ ,
\label{eq-thetaR}
\ea
where $m_{\psi j}$ is an eigenvalue and $\theta_{jk}$ is a rotation angle. Strictly speaking these approximations are only valid for $2\times 2$ matrices when $m_{11}\ll m_{12},m_{21}$, however we have checked, by explicitly performing the calculations in powers of $\lambda_C$, that in all the cases that we will consider with $3\times 3$ matrices they give the correct result to leading order.

In table~\ref{t-soln} we show a set of FN charges and 5D fermion masses that reproduce the neutrino mass matrix of Eq.~(\ref{eq-mnu1}), with:
\be
m_{\nu 3}\simeq v_{H4}\, \frac{g_{5}}{\sqrt{L}} \lambda_C^{20} \ .
\ee
Even though there are many other sets of charges and masses leading to Eq.~(\ref{eq-mnu1}), we have not found any solution with integer charges only, instead at least one of the charges is half-integer. Besides we are interested in configurations with $|c_i|>|c_j|$ for $i<j$, meaning that the lighter generations are more localized towards the UV boundary, and thus have less degree of compositeness, than the heavier ones. Additionally, we are interested in low values of the FN charges, since as we will show in sec.~\ref{sec-LFV}, there are flavor transitions that depend on $p^2$. 
%\ba
%& p_{L1}=2 \ , \qquad p_{L2}=-1 \ , \qquad p_{L3}=-3 \ , \nonumber
%\\
%& p_{N1}=1 \ , \qquad p_{L2}=1 \ , \qquad p_{L3}=1/2 \ , \nonumber
%\\
%& c_{L1}=0.632 \ , \qquad c_{L2}=0.588 \ , \qquad c_{L3}=0.5 \ , \nonumber
%\\
%& c_{N1}=-1.289 \ , \qquad c_{N2}=-1.245 \ , \qquad , c_{N1}=-1.223
%\\
%& n_{L1}=3 \ , \qquad n_{L2}=2 \ , \qquad n_{L3}=0 \ , \nonumber
%\\
%& n_{N1}=18 \ , \qquad n_{N2}=17 \ , \qquad n_{N3}=33/2 \ , \nonumber
%\ea

\begin{table}[ht]
\centering
\begin{tabular}{|c|c|c|c|c|c|c|}
\hline\rule{0mm}{5mm}
generation & $p_L$ & $p_N$ & $c_{\rm L}$ & $c_N$ & $n_L$ & $n_N$ \\[5pt]
\hline 
1st & 2 & 1 & 0.66 & -1.25 & 4 & 17 \\[3pt]
\hline 
2nd & -1 & 1 & 0.61 & -1.21 & 3 & 16 \\[3pt]
\hline 
3rd & -3 & 1/2 & 0.48 & -1.18 & 1 & 31/2 \\[3pt]
\hline 
\end{tabular}
\caption{Benchmark point with FN charges and 5D fermion masses that reproduce the neutrino mass matrix of Eq.~(\ref{eq-mnu1}).}
\label{t-soln}
\end{table}

The 5D masses $c_N$ of table~\ref{t-soln} strongly localize $\nu_R$ of all the generations towards the UV boundary, with values that are similar but not identical. The factor giving the largest suppression of neutrino masses is $F(1+2c_\nu)$. The 5D masses $c_{\rm L}$ generate a moderate suppression for the first and second generations, whereas the Left-handed third generation is only suppressed by $\sim\sqrt{\log(z_1/z_0)}$.

By making use of Eqs.~(\ref{eq-thetaL}) and (\ref{eq-thetaR}) one can estimate the mixing angles of the neutrino sector, obtaining the values of Eq.~(\ref{eq-unu}) for the Left-handed fermions, whereas the Right-handed angles are:
\be
\theta^{\nu_R}_{12}\sim \lambda_C\ ,\qquad \theta^{\nu_R}_{23}\sim \lambda_C\ ,\qquad \theta^{\nu_R}_{13}\sim \lambda_C^2 \ .
\ee

Table~\ref{t-sole} shows a set of benchmark (BM) points for the charged leptons, defined by FN charges and 5D masses. Together with the parameters of the doublets in Table~\ref{t-soln}, these points reproduce the charged-lepton masses and lead to suppressed mixing angles $\theta_{E_{L,R}}^{jk}$. The BMs have different values of $p_{E3}$ and $n_{E3}$, with BM1 having $p_{E3}=-n_{E3}=-3$, and similarly for the subsequent BMs. As we will show in the following sections, some BM points exhibit a parametric suppression of the most constraining flavor observables compared with APC.
\begin{table}[ht]
\centering
\begin{tabular}{|c|c|c|c|}
\hline\rule{0mm}{5mm}
generation & $p_E$ & $c_E$ & $n_E$ \\[5pt]
\hline 
1st & 1 & -0.66 & 4 \\[3pt]
\hline 
2nd & -1 & -0.61 & 3 \\[3pt]
\hline 
3rd & $-3,-5/2,-2,-3/2$ & $-0.61,-0.58,-0.56,-0.53$ & $3,5/2,2,3/2$ \\[3pt]
\hline 
\end{tabular}
\caption{Benchmark points with FN charges and 5D fermion masses that determine the sector of charged leptons. For the 3rd generation we show four solutions that define four BM points: BM1, BM2, BM3 and BM4.}
\label{t-sole}
\end{table}

For the BMs we obtain:
\be
m_E\sim \frac{g_{5}}{\sqrt{L}}v_{H4} \left(\begin{array}{ccc} \lambda_C^9 & \lambda_C^{10} & \lambda_C^{12-9} \\ \lambda_C^9 & \lambda_C^6 & \lambda_C^{8-5} \\ \lambda_C^9 & \lambda_C^6 & \lambda_C^4 \end{array}\right) \ ,
\label{eq-mE}
\ee
where $12\!-\!9$ ($8\!-\!5$) denotes the values 12, 11, 10, 9 (8, 7, 6, 5) corresponding to BM1--BM4. In Ap.~\ref{ap-diag} we show the diagonalization of these matrices perturbatively in powers of $\lambda_C$.

It is interesting to compare the size of the rotation angles in the charged sector with APC. Considering the Left-Right symmetric limit: $c_{\rm L}^j\simeq c_{E}^j$~\cite{Panico:2015jxa}, the mixing angles in APC are given by $\theta^{jk}\sim\sqrt{m_{e}^j/m_{e}^k}$ for $j<k$. By using these estimates as well as Eqs.~(\ref{eq-thetaL}) and (\ref{eq-thetaR}) for the BM points, we obtain the results of table~\ref{t-thetaij}, expressed in terms of $\log_{\lambda_C}\theta^{jk}$. For all the BM points the mixing angles are suppressed in our model compared with APC, with the only exception of $\theta_{e_L}^{23}$ that, for BM4, is of the same order as in APC. $\theta_{e_L}^{12}$, as well as all $\theta_{e_R}^{jk}$, do not change as we change the BM point, whereas $\theta_{e_L}^{23}$ and $\theta_{e_L}^{13}$ depend on the BM, increasing their size from BM1 to BM4.
\begin{table}[ht]
\centering
\begin{tabular}{|c|c|c|c|c|c|c|}
\hline\rule{0mm}{5mm}
$\log_{\lambda_C}\theta^{jk}$ & $\theta_{e_L}^{12}$ & $\theta_{e_L}^{23}$ & $\theta_{e_L}^{13}$ & $\theta_{e_R}^{12}$ & $\theta_{e_R}^{23}$ & $\theta_{e_R}^{13}$ \\[5pt]
\hline 
APC & 1.7 & 1 & 2.7 & 1.7 & 1 & 2.7 \\[3pt]
\hline\rule{0mm}{4mm}
Model & 4 & $4-1$ & $8-5$ & 3 & 2 & 5 \\[3pt]
\hline 
\end{tabular}
\caption{Mixing angles in the charged sector expressed as $\log_{\lambda_C}\theta^{jk}$ in APC and in the model. The interval $4-1$ ($8-5$) means that for BM1, BM2, BM3, BM4 we obtain: $4,3,2,1$ ($8,7,6,5$).}
\label{t-thetaij}
\end{table}
%
%\begin{table}[ht]
%\centering
%\begin{tabular}{|c|c|c|c|c|c|c|}
%\hline\rule{0mm}{5mm}
%model & $\theta^{e_L}_{12}$ & $\theta^{e_L}_{23}$ & $\theta^{e_L}_{13}$ & $\theta^{e_R}_{12}$ & $\theta^{e_R}_{23}$ & $\theta^{e_R}_{13}$ \\[5pt]
%\hline 
%anarchic PC & $\lambda_C^2$ & $\lambda_C$ & $\lambda_C^{2.6}$ & $\lambda_C^2$ & $\lambda_C$ & $\lambda_C^{2.6}$ \\[3pt]
%\hline\rule{0mm}{4mm}
%FN+PC & $\lambda_C^4$ & $\lambda_C^{8-5}$ & $\lambda_C^{4-1}$ & $\lambda_C^3$ & $\lambda_C^5$ & $\lambda_C^2$ \\[3pt]
%\hline 
%\end{tabular}
%\caption{.}
%\label{t-thetaij1}
%\end{table}

The flavor dependent values of $c$ leads to different KK fermion masses. For the BM points of tables~\ref{t-soln} and~\ref{t-sole}, the mass of the first KK fermion doublet ${\rm L}^{(1)}$ and singlet $E^{(1)}$ is $m_1=(2.46-2.65)/z_1$, depending on $c$, with the lightest value for $|c|=0.48$ and the largest one for $|c|=0.66$, whereas for $N^{(1)}$ we get $m_1=(3.39-3.49)/z_1$.

\subsection{Interactions with resonances in the mass basis}
In the next section, we will study the contributions to flavor observables from KK states. Since they receive corrections from the KK mixing, one has to consider the diagonalization of the charged lepton mass matrix. 
For simplicity we will consider the case of one generation only, the results can be extended to three generations straightforwardly. The mass matrix of the charged leptons of Eq.~(\ref{eq-zeta}) is given by:
\be
M_E=v_{H4}y_E+{\rm diagonal}(0,m_1^{\rm L},m_1^E) \ ,
\label{eq-me1}
\ee
where $y_E$ is defined in Eq.~(\ref{eq-ye}) and the second term is a diagonal $3\times 3$ matrix containing the masses of the first level of charged KK states fermions. $M_E$ can be diagonalized in two steps~\cite{Agashe:2006iy}~\footnote{Notice that in Ref.~\cite{Agashe:2006iy} the basis is ordered differently and the mass matrix is taken symmetric.}. ($i$) Taking the limit of vanishing $F$, in which case there is no mixing with the 0-modes and only the lower-right $2\times 2$ block of $M_E$ is non-trivial. This block can be diagonalized exactly with a bi-unitary transformation. ($ii$) Restoring the $F$ dependence to $M_E$ and performing a perturbative diagonalization of the new mass matrix expanding in powers of $F$, as well as in powers of $g_5 v_{H4}/\sqrt{L}m_{\rm KK}$.
We present a summary of results of the mass matrix diagonalization in Ap.~\ref{ap-diag}.

Once the rotation matrices are computed, it is straightforward to obtain the couplings with the Higgs field and with the vector resonances. We will use a tilde to denote masses and couplings of fermions in this basis: $\tilde m_\beta$, $\tilde y_{\alpha\beta}$ and $\tilde g_{\alpha\beta}$. The results can be found in Ap.~\ref{ap-diag}.

%Diagonalization of the fermion mass matrix in the KK space (as if there was just one SM flavor): eigenvalues and rotation matrices.

%Couplings in the mass basis: $h f_{SM} f_{KK}$, $V_{KK}f_{SM}f_{KK}$. Approximate expressions expanding in powers of $g_*v/m_{KK}$ and $F(c)$.

\section{Lepton flavor violation}
\label{sec-LFV}
Now we study CP and LFV in our model~\cite{Goertz:2021ndf,Goertz:2021xlx}. Following Ref.~\cite{Feruglio:2015gka}, we consider dimension six vector and dipole operators, respectively generated at tree and one-loop level. Dimension six scalar and contact operators give subdominant contributions in our model. 

Since we are considering massive neutrinos, we will include the effect of neutrino Yukawa couplings in dipole operators, that give contributions by the virtual-exchange of a charged Higgs. The neutrino Yukawa coupling involving one KK neutrino have a flavor structure determined by $\alpha_N^{jk}$, different from the charged lepton one that depends on $\alpha_E^{jk}$.

First we present the dimension six operators and match their Wilson coefficients to our model. Then we compare the prediction with the case of APC, in this case we will show the results up to factors of ${\cal O}(1)$. We assume maximal CP violating phases.

\subsection{Low energy effective theory for lepton flavor violation}
We extend the SM with dimension six operators made of SM fields:
\be
{\cal L}_{\rm eff}= \sum_j \hat C_j {\cal Q}_j \ .
\label{eq-leff6}
\ee
The Wilson coefficients have flavor indices, $\hat C_j^{kl}$, that for the charged leptons take the values $e,\mu,\tau$. A complete set of operators can be found, for example, in Ref.~\cite{Feruglio:2015gka}. We follow the presentation of that reference. 

For this work we are interested in dipole operators
\ba
& {\cal Q}_{eW}=\bar \ell \sigma^{\mu\nu} e \tau^a W_{\mu\nu}^a H \ , \nonumber\\
& {\cal Q}_{eB}=\bar \ell \sigma^{\mu\nu} e B_{\mu\nu} H \ ,
\ea
from which one can obtain the electromagnetic dipole operator
\be
{\cal Q}_{e\gamma}=\cos\theta_{\rm W} {\cal Q}_{eB} - \sin\theta_{\rm W}{\cal Q}_{eW^3} \ .
\ee
We also consider operators quadratic in the Higgs as well as in the lepton fields, also denoted as vector operators:
\ba
& {\cal Q}_{H\ell}^{(1)}=(H^\dagger i\smalloverleftrightarrow{D}_\mu H)(\bar \ell \gamma^\mu \ell) \ , \nonumber\\
& {\cal Q}_{H\ell}^{(3)}=(H^\dagger i\smalloverleftrightarrow{D}_\mu^a H)(\bar \ell \gamma^\mu \tau^a \ell) \ , \nonumber\\
& {\cal Q}_{He}=(H^\dagger i\smalloverleftrightarrow{D}_\mu H)(\bar e \gamma^\mu e) \ ,
\label{eq-op6h2l2}
\ea
that after EWSB modify the $Z$ couplings. 
%Effects from contact four fermion operators arising from the integration of KK vectors are subdominant in our model~\cite{Agashe:2006iy}.
%Four fermion operators, as $\ell\ell\ell\ell$ and $\ell\ell qq$, contribute to muon and tau decays to three charged leptons and $\mu-e$ conversion in nuclei, respectively. However their contributions are subleading compared with ${\cal Q}_{H\ell}$ and ${\cal Q}_{He}$.

Matching the Wilson coefficient of the electromagnetic dipole operator we obtain:
%\be
%\hat C _{e\gamma}\simeq \hat C _{e\gamma}^h+\hat C _{e\gamma}^{V^{(1)}} \ .
%\ee
%  the Higgs and vector contributions are given by
\ba
& \hat C _{e\gamma}^h\approx -\frac{e}{(4\pi)^2v}\sum_{\beta=2,3}\tilde y_{1\beta}\frac{1}{\tilde m_\beta}\tilde y_{\beta 1} \ ,
\nonumber\\
& \hat C _{e\gamma}^{Z^{(1)}}\approx -\frac{e}{(4\pi)^2v}\sum_{\beta=2,3}\tilde g_{L,1\beta}\tilde m_\beta \tilde g_{R,\beta 1}J(\tilde m_\beta^2,m_{Z^{(1)}}^2) \ ,
\label{eq-Wdip}
\ea
where 
\be
J(x,y)=\dfrac{y\log(y/x)+x-y}{(x-y)}\ 
%\underset{\raisebox{1.2ex}{\substack{\longrightarrow \\ a\to b}}{}}
%\mathrel{\vcenter{\hbox{$\xrightarrow[{\scriptstyle m_\beta\to m_1}]{\phantom{X}}$}}}
\rightarrow
\ \frac{1}{2x}\ ,
\ee
with the r.h.s corresponding to the limit $y\to x$. In our estimates and numerical results of secs.~\ref{sec-estWC} and~\ref{sec-numbers2} we will include the contribution to dipole operators arising from the charged Higgs, that depends on $y_{N}$ and is misaligned with the neutral one, as well as the contribution from $F^{(1)}$, that depends on the FN charges of the fermions.

Matching the Wilson coefficient of vector operators and denoting as $Q^Z_{\beta}$ the $Z$-charge of the fermion $\beta$, we get:
\ba
& \hat C _{H\ell}^{(1)}\approx \sum_{\beta=2,3}\tilde y_{1\beta}\frac{1}{\tilde m_\beta}Q^Z_{\beta}\frac{1}{\tilde m_\beta}\tilde y_{1\beta}^\dagger + \frac{g^{hh}_{(0n)}\tilde g_{L,11}}{m_{Z^{(1)}}^2}\ ,
\nonumber\\
& \hat C _{H\ell}^{(3)}=0\ ,
\nonumber\\
& \hat C _{He}\approx \sum_{\beta=2,3}\tilde y_{\beta 1}^\dagger\frac{1}{\tilde m_\beta}Q^Z_{\beta}\frac{1}{\tilde m_\beta}\tilde y_{\beta 1} + \frac{g^{hh}_{(0n)}\tilde g_{R,11}}{m_{Z^{(1)}}^2}\ .
\label{eq-Wvec}
\ea

To analyse the flavor structure of the Wilson coefficients one can expand them to leading order in $F(c)$, $g_5v_{H4}/\sqrt{L}m_{\rm KK}$ and $g_{\rm SM}\sqrt{L}/g_5$, obtaining:
\ba
& \hat C _{e\gamma}^h\approx \frac{3e}{(4\pi)^2}y_{13}\frac{1}{m_3}y_{32}\frac{1}{m_2}y_{2 1} \ ,
\nonumber\\
& \hat C _{e\gamma}^{Z^{(1)}}\approx \frac{e}{(4\pi)^2}\frac{1}{2m^2_{Z^{(1)}}}\left(y_{13}g_{L33}g_{R,13}-g_{L,12}g_{R,22}y_{21}+g_{L,12}y_{23}g_{R,13}\right)\ ,
\nonumber\\
& \hat C _{H\ell}^{(1)}\approx y_{13}\frac{1}{2m_3^2}y_{13}^\dagger + \frac{g^{hh}_{(0n)}\tilde g_{L,11}}{m_{Z^{(1)}}^2}\ ,
\nonumber\\
& \hat C _{He}\approx -y_{21}^\dagger\frac{1}{2m_2^2}y_{21} + \frac{g^{hh}_{(0n)}\tilde g_{R,11}}{m_{Z^{(1)}}^2}\ .
\label{eq-WCs}
\ea
The flavor structure of the couplings is given in Eqs.~(\ref{eq-yejk}) and~(\ref{eq-gni}).

Finally, one has to rotate to the mass basis in the generation space~\ref{ap-diag}:
\ba
C_{e\gamma}=U_{e_L}\hat C _{e\gamma}U_{e_R}^\dagger, \ \qquad 
C_{H\ell}=U_{e_L} \hat C _{H\ell} U_{e_L}^\dagger\ , \qquad 
C_{He}=U_{e_R} \hat C _{He} U_{e_R}^\dagger\ . \qquad 
\ea

\subsection{Estimates of the Wilson coefficients and comparison with APC}
\label{sec-estWC}
In this subsection, we compare the Wilson coefficients predicted by our model with those of APC in Ref.~\cite{Feruglio:2015gka}. As in sec.~\ref{sec-lightleptons}, we consider the Left–Right symmetric limit for APC, in which case the degree of compositeness of the chiral charged leptons is directly determined by their masses~\cite{Panico:2015jxa}. Using Eq.~(\ref{eq-emass}), the corresponding Wilson coefficients can then be expanded in powers of $\lambda_C$.
Also in our model, with $U_{e_{L,R}}$, $F(1\mp c)$ and $\delta$ parameterized in powers of $\lambda_C$, the Wilson coefficients admit an expansion in the same parameter. 

We obtain the following result for the electron EDM:
\be
C_{e\gamma}^{e}\sim (C_{e\gamma}^{e})|_{\rm APC} \lambda_C^2\ , \qquad
\label{eq-CeEDM}
\ee
dominated by both, the neutral and the charged Higgs contributions, for the four BM points of table~\ref{t-sole}. 
Earlier bounds on the electron EDM implied that in APC $m_{KK}/(g_5/\sqrt{L}) \gtrsim 30\,{\rm TeV}$~\cite{ACME:2013pal,Cairncross:2017fip}, 
but the latest results~\cite{ACME:2018yjb,Roussy:2022cmp} significantly strengthen this constraint to $100\text{--}175\,{\rm TeV}$. 
Although the suppression in Eq.~(\ref{eq-CeEDM}) lowers the bound on the NP scale by a factor $\lambda_C$, obtaining $m_{KK}/(g_5/\sqrt{L}) \gtrsim 22-39\,{\rm TeV}$, reaching the TeV scale requires $C_{e\gamma}^{e}\sim (C_{e\gamma}^{e})|_{\rm APC}\,\lambda_C^6$.

For the flavor violating dipole operators we get:
\ba
C_{e\gamma}^{\mu e}\sim (C_{e\gamma}^{\mu e})|_{\rm APC} \lambda_C^{3/2}
\ , \qquad
C_{e\gamma}^{\tau\mu}\sim (C_{e\gamma}^{\tau\mu})|_{\rm APC} \lambda_C^{1,1,1,0}
\ , \qquad
C_{e\gamma}^{\tau e}\sim (C_{e\gamma}^{\tau e})|_{\rm APC} \lambda_C^{5/2}
\ .
\label{eq-Cdipoles}
\ea
In $C_{e\gamma}^{\mu e}$ the dominant contribution is generated by $F^{(1)}$ exchange in the operator with the chiral structure $\bar \mu_L\sigma^{\nu\rho}e_R$, whereas the operator with flipped chiralities is suppressed by and extra factor $\lambda_C$. 
Experimental bounds require in APC $m_{KK}/(g_5/\sqrt{L}) \gtrsim 32\,{\rm TeV}$~\cite{ParticleDataGroup:2024cfk}, the suppression in the first line of Eq.~(\ref{eq-Cdipoles}) lowers the bound to $\sim 10\,{\rm TeV}$.
In $C_{e\gamma}^{\tau\mu}$ the dominant contribution for the first three BM points is dominated by neutral Higgs and $F^{(1)}$ virtual exchange in the operator with the chiral structure $\bar \tau_L\sigma^{\nu\rho}\mu_R$ and is suppressed by $\lambda_C$ compared with APC. For BM4 the dominant operator is $\bar \mu_L\sigma^{\nu\rho}\tau_R$, all the contributions are of the same order and there is no suppression. 
Finally, $C_{e\gamma}^{\tau e}$ is dominated by $F^{(1)}$ exchange in the operator with the chiral structure $\bar \tau_L\sigma^{\nu\rho}e_R$ for all the BM points, with a suppression factor $\lambda_C^{5/2}$ compared with APC. For BM4 there are contributions of the same order from $F^{(1)}$ in the chiral flipped operator and from Higgs exchange.

For the vector operators in $\mu e$ transitions we obtain the following results:
\be
C_{H\ell}^{(1)\mu e}\sim (C_{H\ell}^{(1)\mu e})|_{\rm APC} \lambda_C^{5/2}\ , 
\qquad
C_{He}^{\mu e}\sim (C_{He}^{\mu e})|_{\rm APC} \lambda_C^{3/2}\ . 
\label{eq-Ovectormue}
\ee
For $C_{H\ell}^{(1)\mu e}$ both terms of Eq.~(\ref{eq-WCs}) result of the same order in powers of $\lambda_C$, whereas for $C_{He}^{\mu e}$ the second one dominates. The suppression compared with APC relaxes the bound on $m_{\rm KK}/(g_5/\sqrt{L})^{1/2}$, from $3\,{\rm TeV}$~\cite{Feruglio:2015gka,ParticleDataGroup:2024cfk} to $0.5\,{\rm TeV}$.

For $\tau\mu$ vector operators we get:
\be
C_{H\ell}^{(1)\tau\mu}\sim (C_{H\ell}^{(1)\tau\mu})|_{\rm APC} \lambda_C^{1,1,-1,-1}\ , 
\qquad
C_{He}^{\tau\mu}\sim (C_{He}^{\tau\mu})|_{\rm APC} \lambda_C^{3,2,1,0}\ ,
\label{eq-Ovectortaumu}
\ee
where the list of exponents refers to the list of BM points, from 1 to 4. For $C_{H\ell}^{(1)\tau\mu}$ the second term of Eq.~(\ref{eq-WCs}) dominates, whereas for $C_{He}^{\tau\mu}$ both are of the same order. The present bounds on decays mediated by this operators do not give relevant constraints in the model.

Finally, for $\tau e$ vector operators we get:
\be
C_{H\ell}^{(1)\tau e}\sim (C_{H\ell}^{(1)\tau e})|_{\rm APC} \lambda_C^{7/2,7/2,5/2,5/2}\ , 
\qquad
C_{He}^{\tau e}\sim (C_{He}^{\tau e})|_{\rm APC} \lambda_C^{9/2,7/2,5/2,3/2}\ ,
\label{eq-Ovectortaue}
\ee
where again the list of exponents correspond to the BM points, from 1 to 4. In all the cases the second term of Eq.~(\ref{eq-WCs}) dominates, except for BM1 and 2 in $C_{H\ell}^{(1)\mu e}$, where the first term is of the same order in powers of $\lambda_C$.

\section{Numerical results}
\label{sec-numbers}
In this section, we present the numerical results obtained within the model for BM1.
The 5D parameters are specified as follows. The 5D fermion masses are given in table~\ref{t-soln}, together with the BM1 point of the charged-lepton sector defined in table~\ref{t-sole}. The 5D coupling in the fermion sector is fixed to $g_5/\sqrt{L}=3$. For the scalar fields we set $\beta_\Phi=\beta_H=0$. In addition, we choose $\delta=\sqrt{\Lambda_C}$, $z_0=10^{-15}z_1$, and $z_1=1/{\rm TeV}$.
A random scan is then performed over the ${\cal O}(1)$ parameters of the model, namely the coefficients $x_N^{jk}$ and $x_E^{jk}$ of the 5D Yukawa couplings in Eq.~(\ref{eq-Ly}), restricted to the range $|x_{N,E}^{jk}|\in(1/3,3)$.

We do not aim to fit the lepton observables with precision, instead we we will show that the chosen scenario can naturally reproduce masses and mixing angles close to the observed values. Given a set of parameters that is closed to the measured values, by small adjustments of the parameters one can reproduce them with precision.

We present first the predictions for the masses and mixing angles of the light leptons, and then we show the results for the Wilson coefficients of the flavor- and CP-violating operators.

\subsection{Lepton masses and mixing angles}
\label{sec-numbers1}
\begin{figure}[t]
\centering
\includegraphics[width=0.34\textwidth]{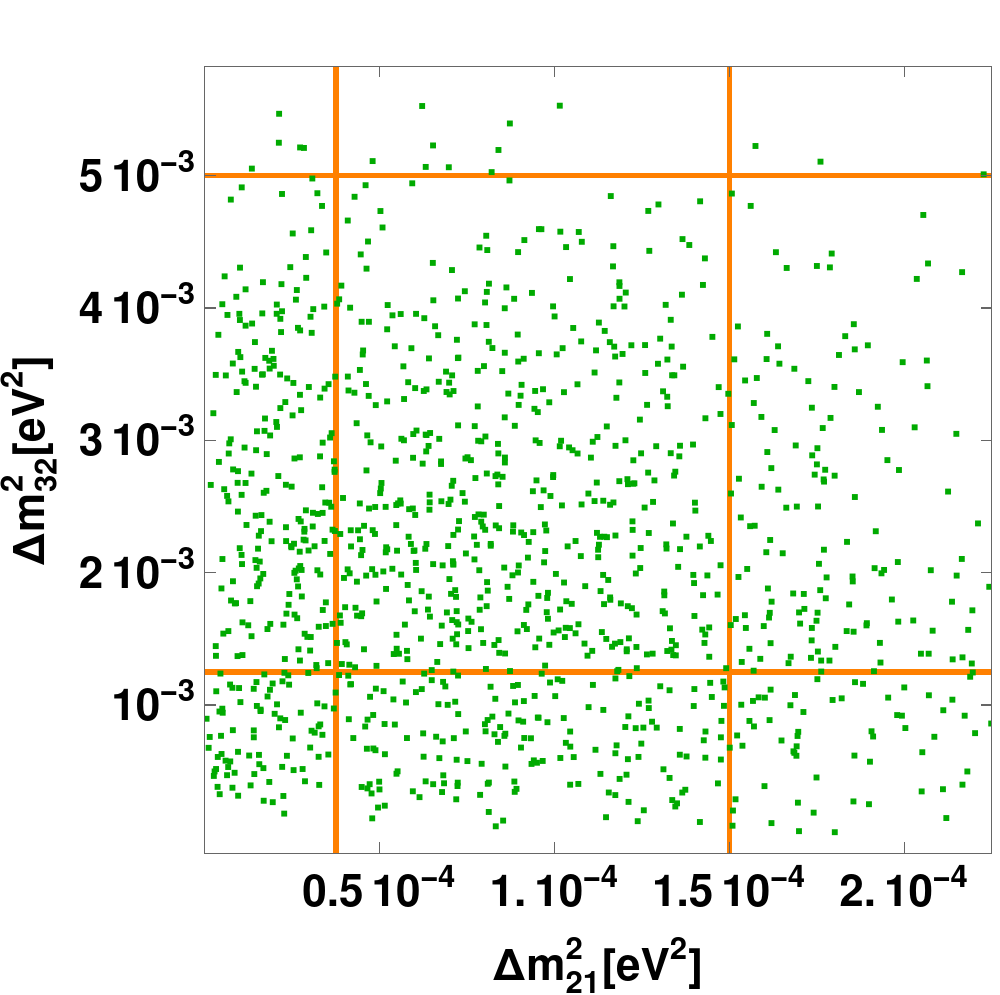}
\includegraphics[width=0.32\textwidth]{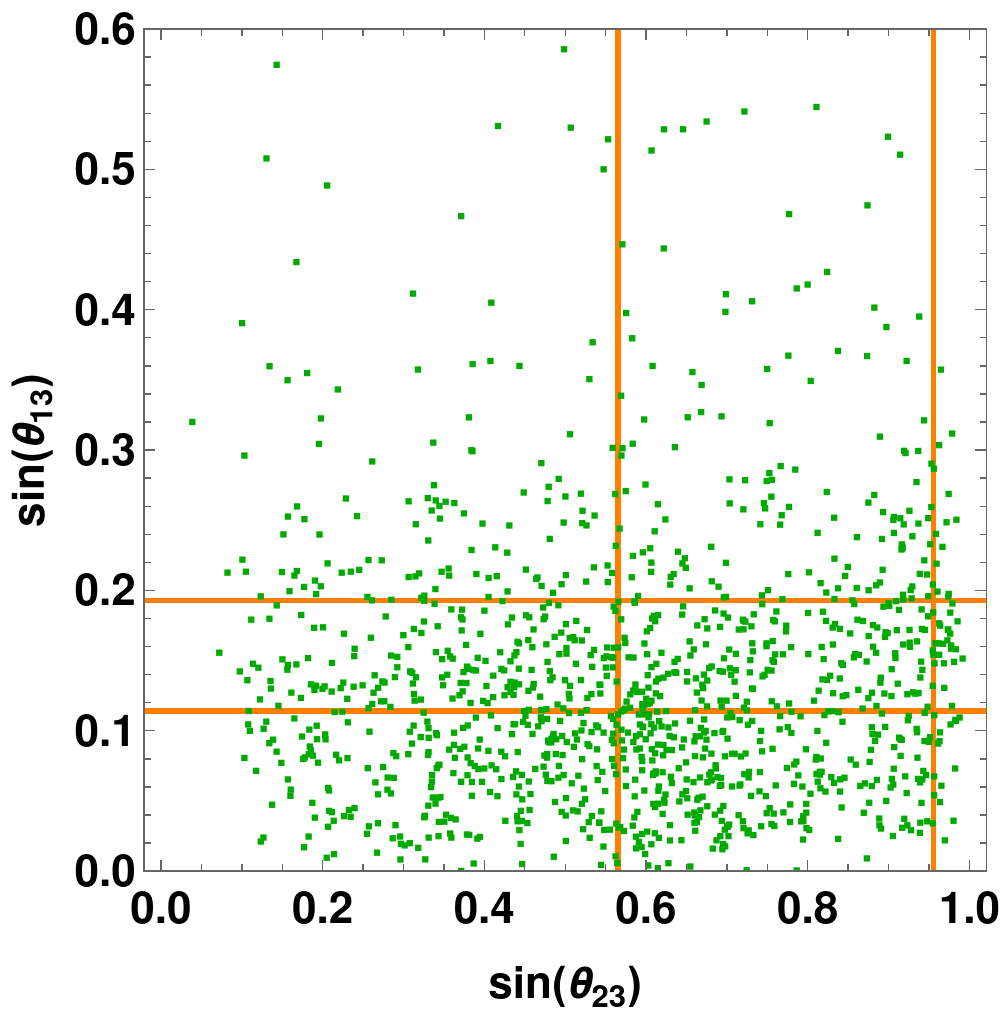}
\includegraphics[width=0.32\textwidth]{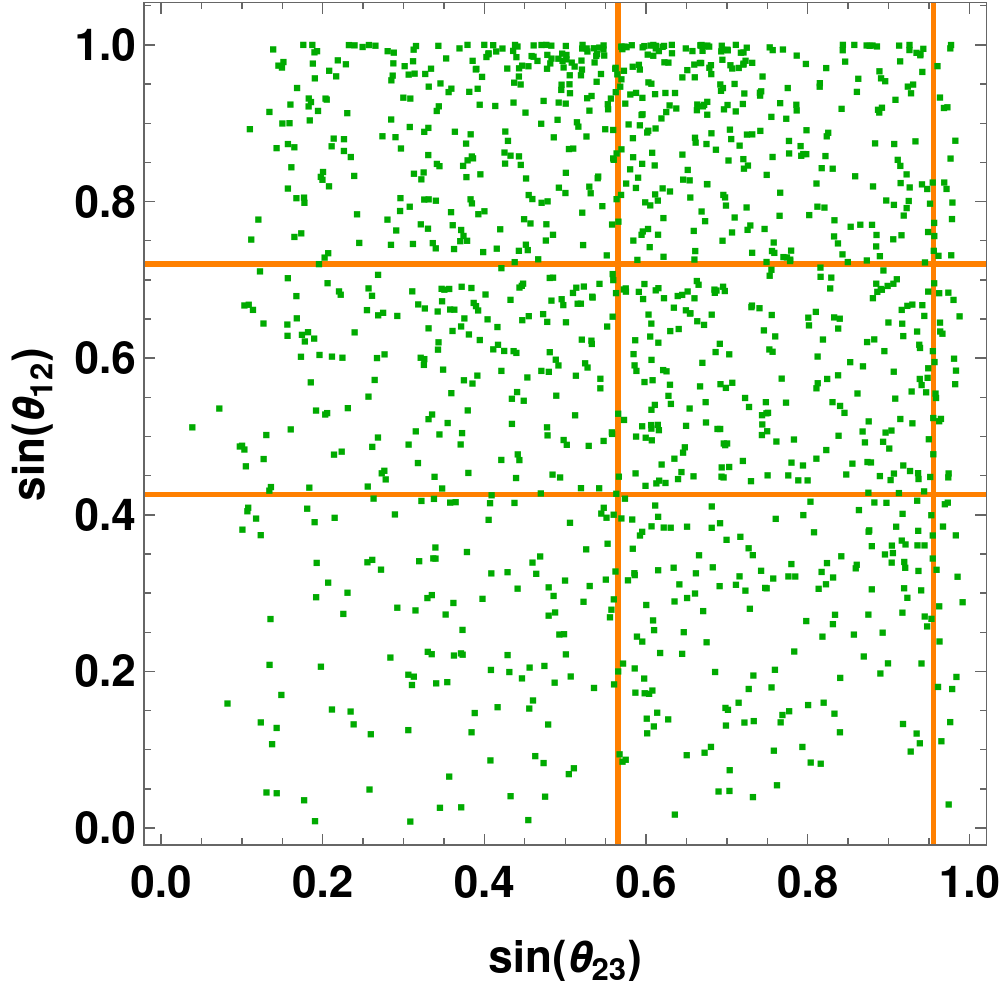}
\caption{Numerical predictions of the neutrino observables with BM1 of the model, performing a random scan over the Yukawa couplings. On the left we show the mass-squared differences and in the centre and in the right the mixing angles of the PMNS matrix.}
\label{fig-mnu}
\end{figure}
In Fig.~\ref{fig-mnu} we show the results for the neutrino mass squared differences and the mixing angles of the PMNS matrix. The plot on the left displays $\Delta m_{ij}^2$ with the orange lines corresponding to departures by a factor $1.5$ from the measured values, whereas the plots on the centre and on the right show $\sin\theta_{ij}$ with the orange line corresponding to a tolerance factor $1.3$. From the random scan, the mean value and standard deviation of the mass-squared differences are: $\Delta m^2_{21}=(9.7\pm 6.7)\times 10^{-5}\,{\rm eV}^{2}$ and $\Delta m^2_{32}=(2.2\pm 1.2)\times 10^{-3}\,{\rm eV}^{2}$, while the mixing angles are: $|\sin\theta_{12}|\approx 0.64\pm 0.27$, $|\sin\theta_{23}|\approx 0.58\pm 0.24$ and $|\sin\theta_{13}|\approx 0.14\pm 0.09$.

\begin{figure}[h]
\centering
\includegraphics[width=0.33\textwidth]{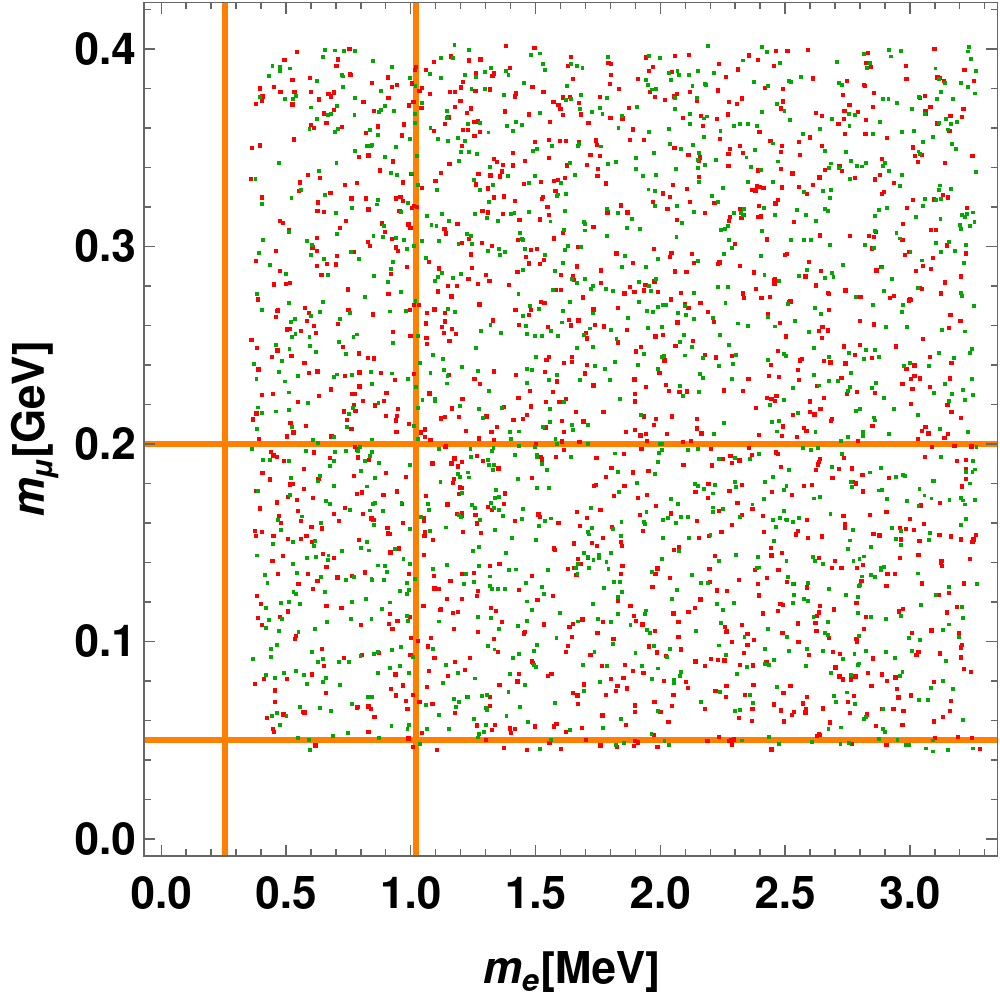}
\hspace{10mm}
\includegraphics[width=0.32\textwidth]{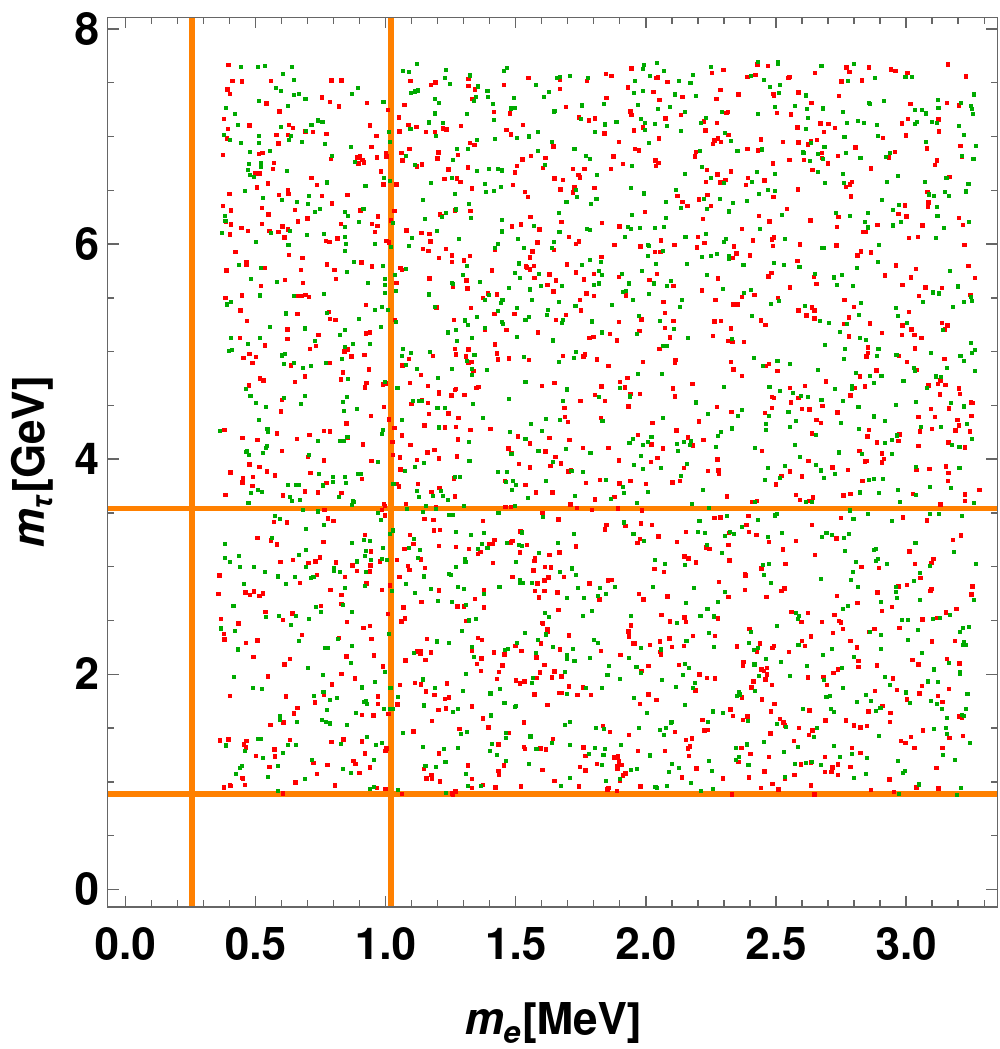}
\caption{Numerical predictions of the masses of the charged leptons with BM1 of the model, performing a random scan over the Yukawa couplings. Orange lines show the analytical estimates.}
\label{fig-me}
\end{figure}

In Fig.~\ref{fig-me} we present the results for the charged lepton masses. 
The green points correspond to randomly generated samples, while the red points show a subset of the scan that satisfies the neutrino-sector constraints, allowing deviations within a factor of 2 for the measured values of $\Delta m_{kj}^2$ and within a factor of 1.3 for $\sin\theta_{jk}$.
The orange lines indicate $m_i/2$ and $2m_i$, with $i=e,\mu,\tau$ denoting the electron, muon, and tau masses, respectively. 
As expected, no structure is observed for the red points, since the neutrino masses and mixings depend on $x_{N}^{jk}$, while the charged ones depend on $x_{E}^{jk}$. Although the masses turn out somewhat larger than their physical values on average, a more accurate agreement can be achieved by selecting slightly different values of $c_{E}^j$, for example by decreasing $n_E^j$ by $1/2$, provided we allow for non-integer $n_E^j$.\footnote{It is not necessary to restrict $n_{\psi}^j$ to integer values, this choice was adopted to simplify the calculation of the eigenvalues and eigenvectors of the mass matrix, expanded in series in powers of $\lambda_C$.}

\subsection{Numerical results for the Wilson coefficients}
\label{sec-numbers2}
%In Fig.~\ref{fig-dipoles} we report the numerical results predicted by the model for the Wilson coeficient of dipole operators, we present them normalized to the prediction in APC asuming the Left-Right symmetric limit. The orange lines correspond the estimates for BM1 given in Eqs.~(\ref{eq-CeEDM}-\ref{eq-Ovectortaue}), the gray lines correspond to $C=C|_{\rm APC}$, the green and red points are defined as in the previous subsection, the former ones from random samples and the later ones selecting from those samples points in agreement with the measured values. No difference between the distributions of both sets can be appreciated in the plots. The cusps on the upper right region of the point distributions are obtained for maximal values of the Yukawa couplings considered in the scan. On the upper left pannel we show for the electron EDM the contributions from the neutral and charged Higgs on the $x$ and $y$ axis, respectively. The estimate of Eq.~(\ref{eq-CeEDM}) underestimates the contribution from the neutral Higgs for some points, but there are also points where $C^e_{e\gamma}$ is an order of magnitude smaller than the estimate, presumably arising from accidental numerical cancellations. 	

In this subsection we present the model's numerical predictions in the case of BM1 for the Wilson coefficients, normalized to the prediction of APC in the Left-Right symmetric limit: $|C/C_{\rm APC}|$. 
\begin{figure}[h!]
\centering
\includegraphics[width=0.407\textwidth]{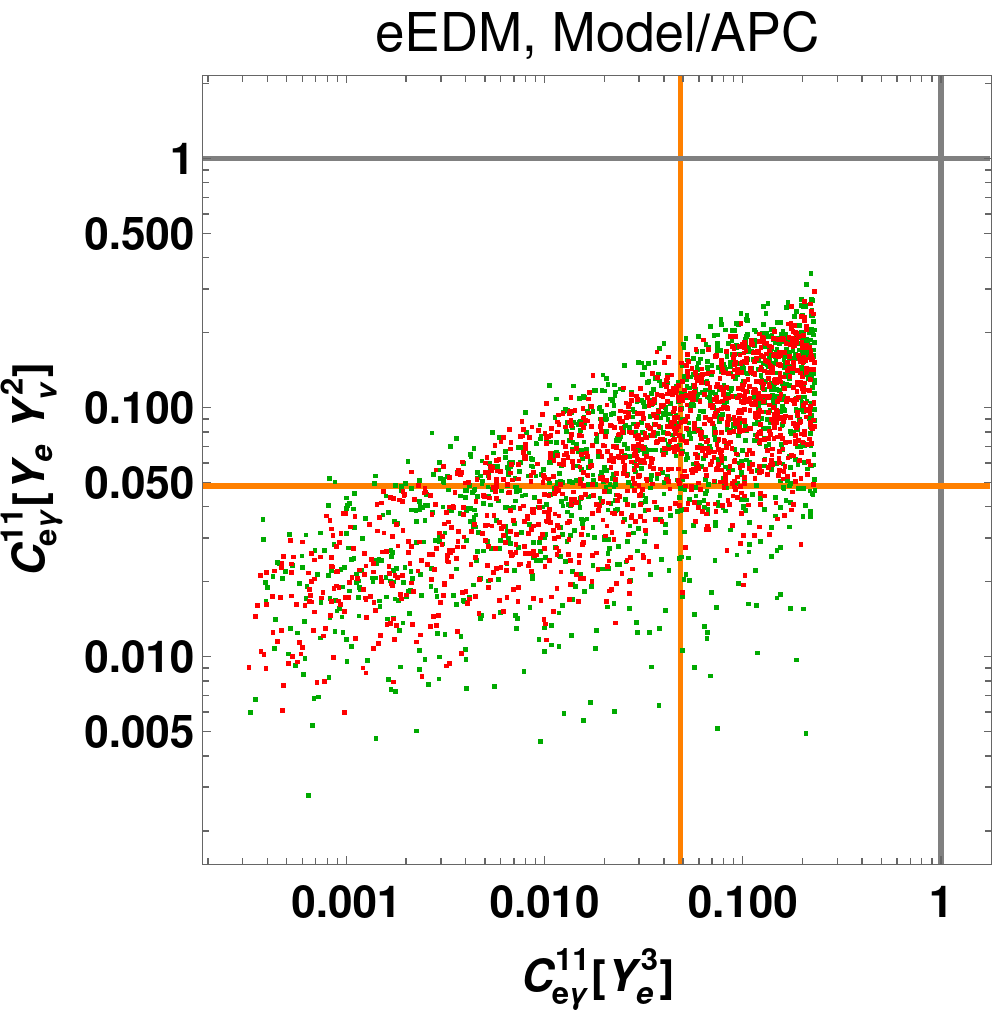}
\hspace{10mm}
\includegraphics[width=0.4\textwidth]{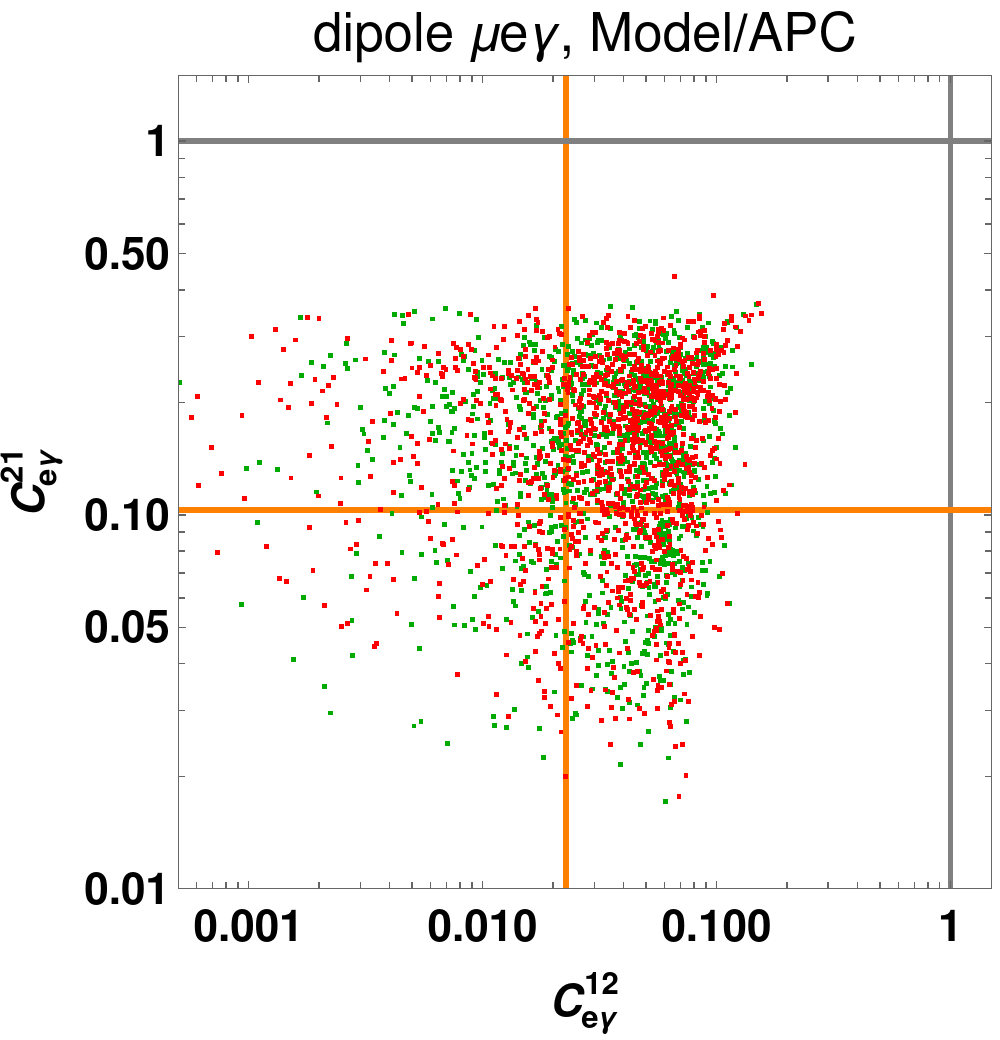}\\
\vspace{10pt}
\includegraphics[width=0.4\textwidth]{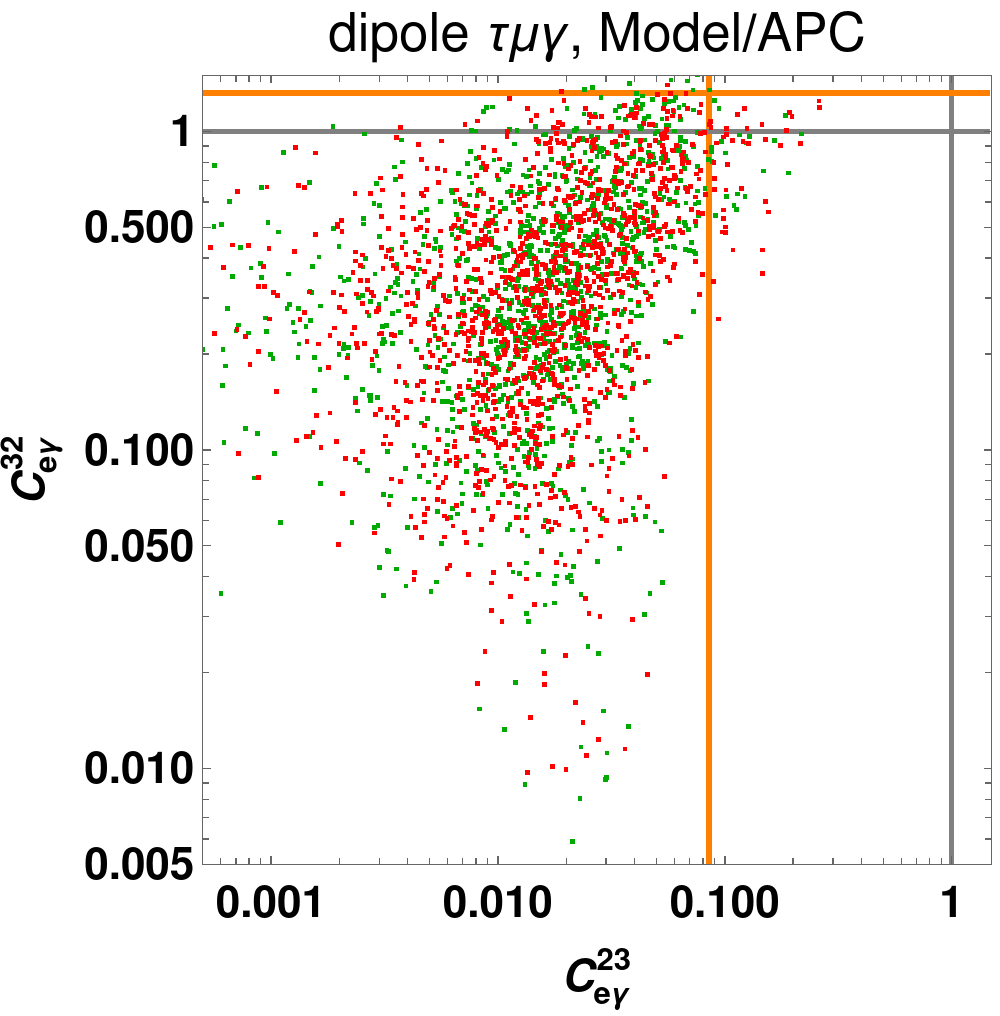}
\hspace{10mm}
\includegraphics[width=0.4\textwidth]{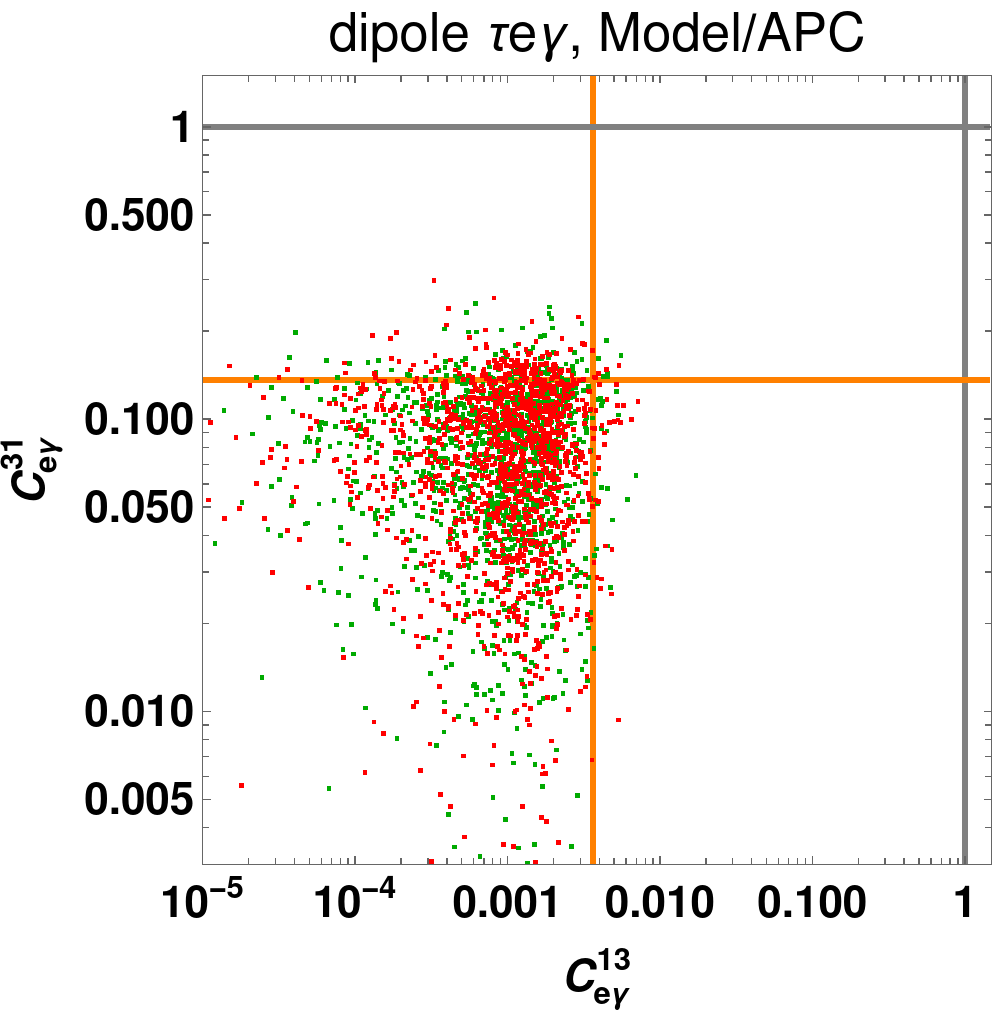}
\caption{Numerical predictions of the Wilson coefficients of dipole operators with BM1 of the model, normalized with respect to APC, performing a random scan over the Yukawa couplings. Orange lines show the analytical estimates. On the top-left panel we show the neutral and charged Higgs contributions to the $e$EDM, on the top-right we show the different chiralities in $\mu\to e\gamma$, as on the bottom-left and bottom-right panels, where we show respectively $\tau\to \mu\gamma$ and $\tau\to e\gamma$.}
\label{fig-dipoles}
\end{figure}

Figure~\ref{fig-dipoles} presents the prediction for dipole operators.
The results from a random parameter scan are shown as green points, with a subset of these points that reproduce the experimental masses and mixing of neutrinos highlighted in red.~\footnote{Selecting from them the points that reproduce the masses of the charged leptons withing corrections by a factor 2 does not give significant changes in the distributions of points.} The distributions of both sets show no appreciable difference. For reference, the analytical estimates (Eqs.~(\ref{eq-CeEDM}-\ref{eq-Ovectortaue})) are indicated by orange lines, with the gray lines corresponding to $C=C|_{\rm APC}$. The cusps visible in the upper-right region of the distributions correspond to the maximum values of the Yukawa couplings allowed in the scan. 

The top-left panel displays the contributions to the electron EDM operator, $C^e_{e\gamma}$, with the neutral Higgs contribution on the $x$-axis and the charged Higgs contribution on the $y$-axis. Compared to the estimate in Eq.~(\ref{eq-CeEDM}), the neutral Higgs contribution is larger for some points, whereas for others the total value of $C^e_{e\gamma}$ is an order of magnitude smaller than the estimate, presumably due to accidental numerical cancellations. The mean value and standard deviation of the points are: $0.066\pm 0.068$ and $0.080\pm 0.058$, in the $x-$ and $y-$axis, respectively, of the same order of magnitude as the estimate $\lambda_C^2\simeq 0.048$.

The top-right panel shows both chiral structures of $C_{e\gamma}^{\mu e}$. Many points cluster slightly above the estimate, within a factor $\sim 3-5$, and a small subset are suppressed by approximately one order of magnitude. The mean value and standard deviation of the coordinates of the points are: $0.043\pm 0.026$ and $0.16\pm 0.08$, of the same size as the respective estimates $\lambda_C^{5/2}\simeq 0.023$ and $\lambda_C^{3/2}\simeq 0.10$.

The bottom-left panel displays $C_{e\gamma}^{\tau\mu}$ for both chiral structures. In this case the orange lines represent analytical estimates that include sizable numerical factors obtained in the full calculation: $C_{e\gamma}^{\tau\mu,23}/(C_{e\gamma}^{\tau\mu,23})_{\rm APC}\sim (p_{E3}p_{\rm L3}-p_{E2}p_{\rm L2})\lambda_C^3\approx 0.085$ and $C_{e\gamma}^{\tau\mu,32}/(C_{e\gamma}^{\tau\mu,32})_{\rm APC}\sim p_{\rm L3}(p_{E3}-p_{E2})\lambda_C\approx 1.32$. These factors derived from the FN charge in the vertices of the one loop diagram explain why the predicted coefficients can reach the APC reference value, even after accounting for the Cabibbo angle suppression. They also show how large charges can increase the LFV dipole. The mean value and standard deviation of the coordinates of the points are somewhat smaller than the estimates: $0.022\pm 0.021$ and $0.40\pm 0.30$.

Finally, the bottom-right panel presents $C_{e\gamma}^{\tau e}$. Including the numerical prefactors determined by the FN charges the orange lines are given by: $C_{e\gamma}^{\tau e,13}/(C_{e\gamma}^{\tau e,13})_{\rm APC}\sim p_{E3}(p_{\rm L3}-p_{\rm L1})\lambda_C^{11/2}\approx 3.6\times 10^{-3}$ and $C_{e\gamma}^{\tau e,31}/(C_{e\gamma}^{\tau e,31})_{\rm APC}\sim p_{\rm L3}(p_{E3}-p_{E1})\lambda_C^{5/2}/2\approx 0.14$. The mean value and standard deviation of the coordinates of the points are a factor 2-3 smaller: $(1.2\pm 0.9)\times 10^{-3}$ and $0.076\pm 0.043$.

\begin{figure}[h!]
\centering
\includegraphics[width=0.4\textwidth]{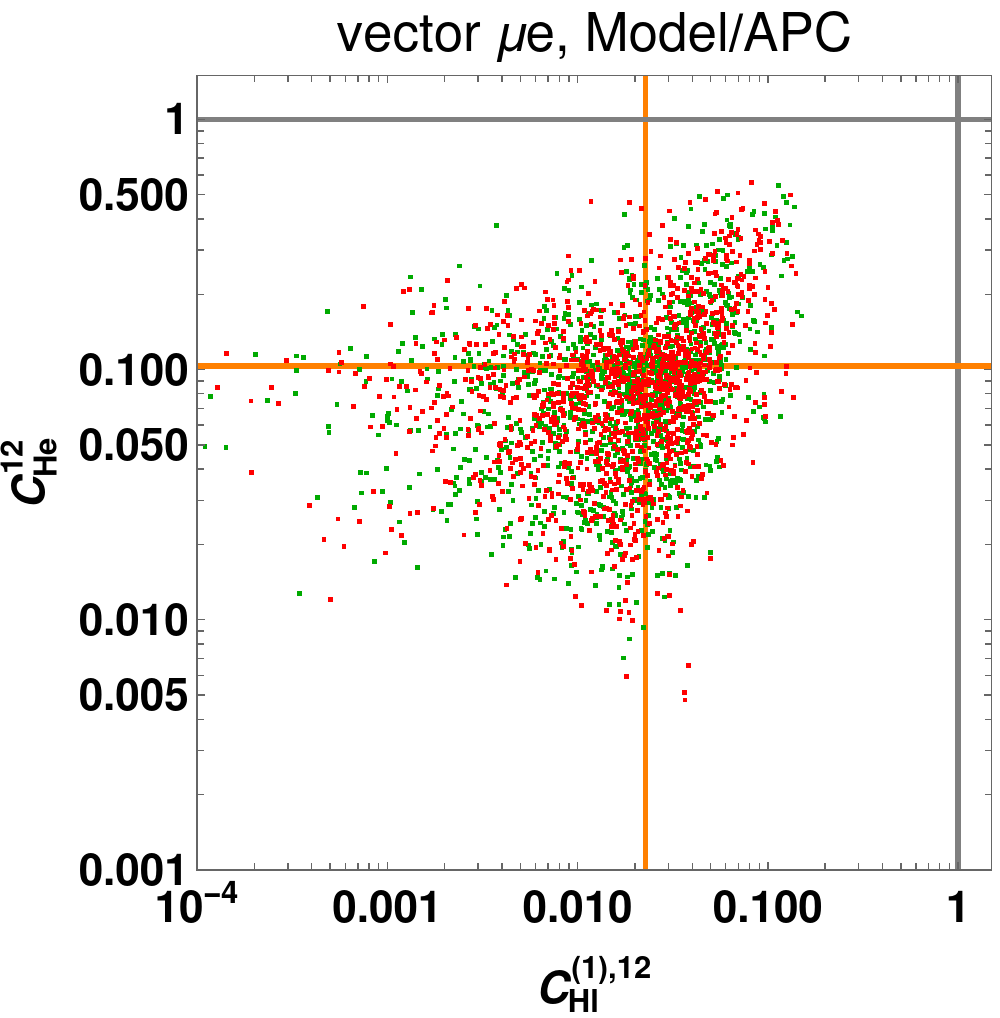}
\hspace{10mm}
\includegraphics[width=0.4\textwidth]{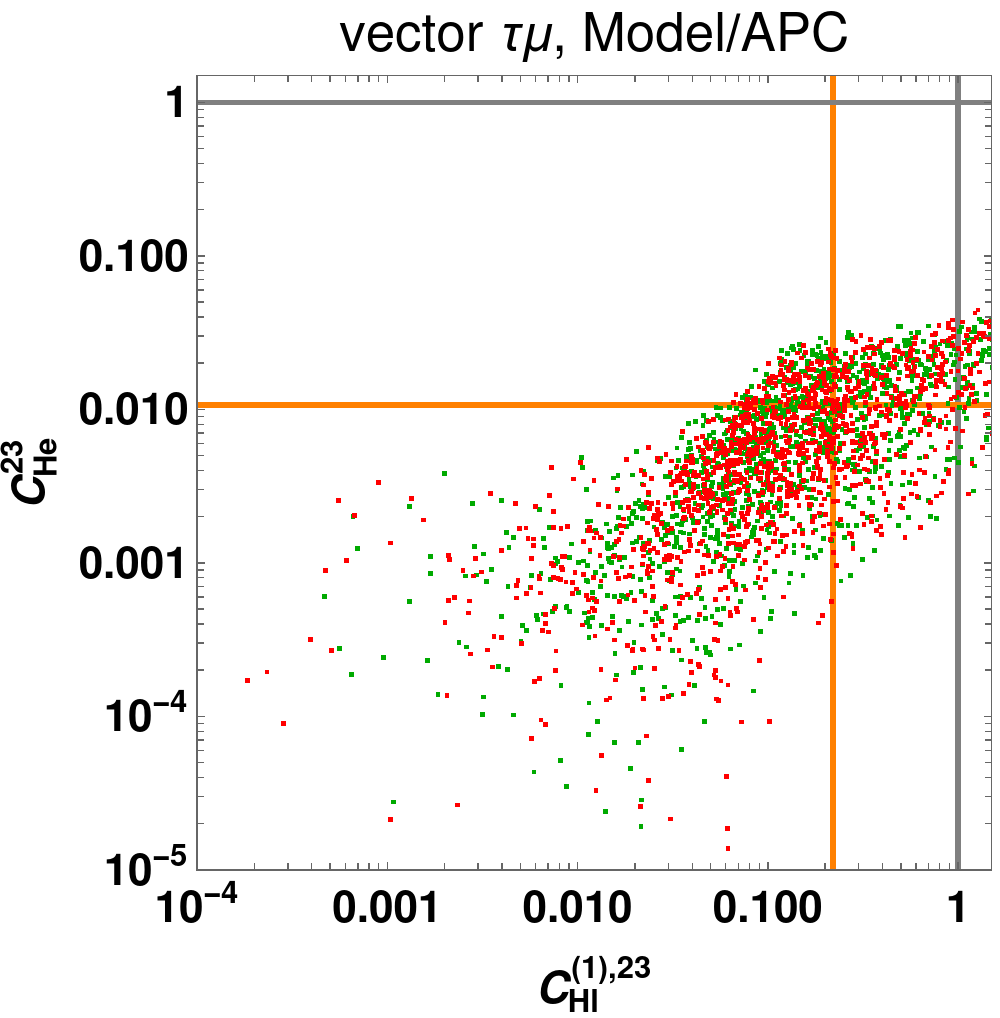}\\
\includegraphics[width=0.4\textwidth]{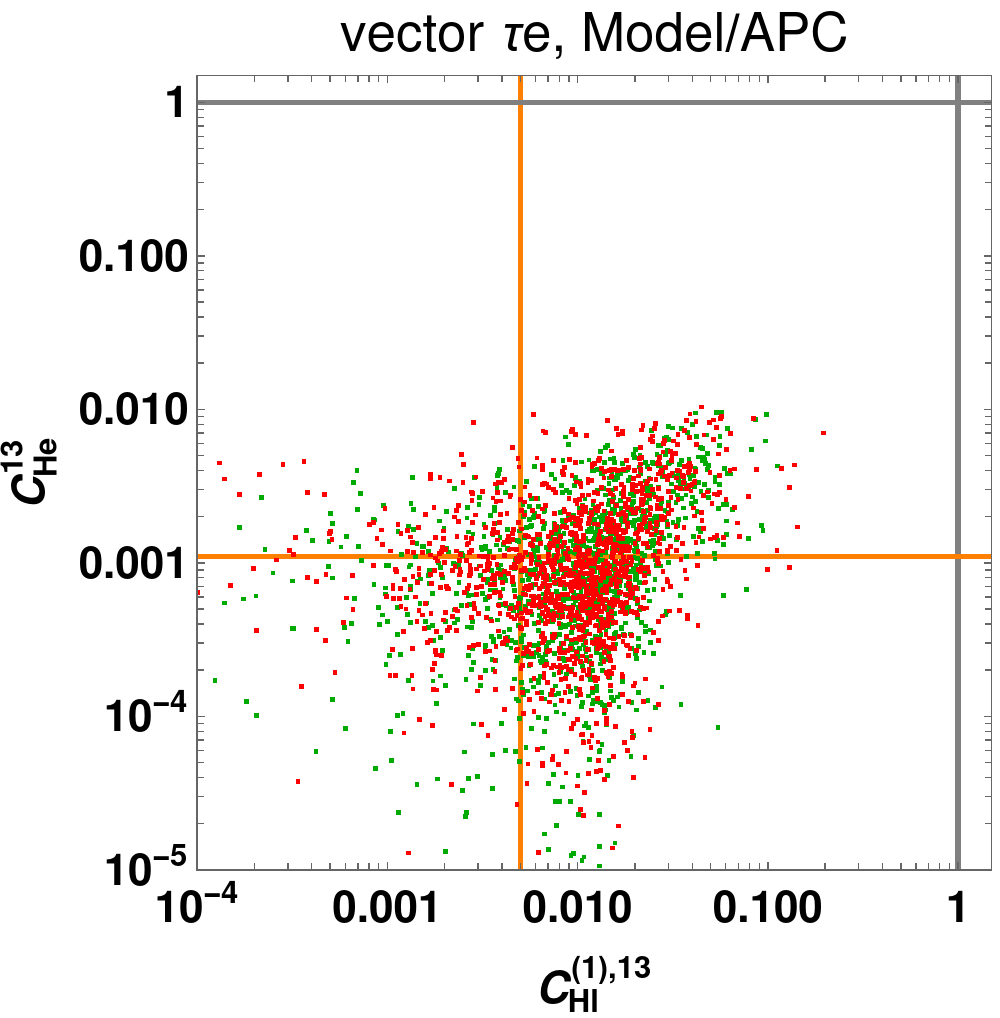}
\caption{Numerical predictions of the Wilson coefficients of vector operators with BM1 of the model, normalized with respect to APC, performing a random scan over the Yukawa couplings. Orange lines show the analytical estimates. On the top-left and top-right panels we show the results for $\mu e$ and $\tau\mu$ operators,respectively, and on the bottom one for $\tau e$.}
\label{fig-vectors}
\end{figure}

Figure~\ref{fig-vectors} presents the Wilson coefficients for the vector operators, normalized to the APC prediction and using the same color scheme as Fig.~\ref{fig-dipoles}. The upper-left panel displays the results for $\mu e$ transitions. The mean value and standard deviation of the coordinates of the points are: $0.025\pm 0.022$ and $0.10\pm 0.08$, closely aligned with the analytical estimates of $\lambda_C^{5/2}\approx 0.023$ and $\lambda_C^{3/2}\approx 0.10$, respectively.

The upper-right panel, corresponding to $\tau\mu$ transitions, yields mean values of $0.33\pm 0.62$ and $0.009\pm 0.010$. These are consistent with the respective estimates of $\lambda_C$ and $\lambda_C^{3}\approx 0.011$, noting the large standard deviation for the first coordinate.

Finally, the bottom panel shows the prediction for $\tau e$ transitions. The the observed values $(1.4\pm 1.3)\times 10^{-2}$ and $(1.4\pm 1.4)\times 10^{-3}$ can be compared with the estimates $\lambda_C^{7/2}\approx 0.4\times 10^{-2}$ and $\lambda_C^{9/2}\approx 1.1\times 10^{-3}$, observing an enhancement by a factor 3 for the first coordinate.

\section{Discussions and conclusions}
\label{sec-conclusions}
One of the main goals of particle physics is to find an explanation of the pattern of flavor of the SM. In the present paper we have presented a 5D model capable of explaining the flavor of the leptonic sector. 
The model incorporates two principal mechanisms: partial compositeness, achieved through the localization of zero modes along the extra dimension, and a bulk horizontal U(1) gauge symmetry, under which the 5D leptons are charged.
We added a 5D scalar field $\Phi$ and an IR boundary potential that triggers EW and U(1) spontaneous symmetry breaking. 
Obtaining a U(1) symmetry breaking parameter of order $\sqrt{\lambda_C}$ requires a rather large 5D coupling in the scalar sector of the theory, dual to $N_{\rm CFT}\simeq 3$.
5D Yukawa interactions are generated by higher dimensional operators with powers of $\Phi/\Lambda$ determined by the fermion charges. In this way, starting from 5D anarchic Yukawa couplings, effective 5D Yukawa interactions with non-trivial flavor patterns are generated, as usual in 4D FN models.    
%With the addition of a 5D scalar field $\Phi$ that spontaneously breaks the horizontal U(1) symmetry, 5D Yukawa interactions are generated by higher dimensional operators with powers of $\Phi/\Lambda$. 
We have shown that under the interplay of both mechanisms it is possible to explain the hierarchy of masses of the light leptons, to be identified with the SM ones, as well as the angles of the PMNS matrix. The neutrinos are realized as Dirac fermions, with normal ordered masses and a hierarchical spectrum, and their mixing angles dominate the PMNS matrix. The strong localization of $\nu^{(0)}_R$ towards the UV boundary suppresses the neutrino masses. 

We have computed the main contributions to CP and LFV processes, that are dominated by vector and dipole operators generated at tree and one loop level, respectively. There is a new contribution to dipole operators compared with APC, obtained by the exchange of U(1) KK resonances that, given the non-universal charges of the fermions, is misaligned with the mass. This contribution is the dominant one for LFV dipoles. We have found configurations of charges and masses of the 5D fermions that are capable of suppressing these processes compared with APC. In particular, the $e$EDM can be suppressed by $\lambda_C^2$, relaxing the bound of APC, but still requiring a compositeness scale $m_*/g_*\gtrsim 22-39\,{\rm TeV}$. The bound from $\mu\to e\gamma$ can be relaxed by a factor ${\cal O}(3)$, leaving the compositeness scale in the ballpark of $10\,{\rm TeV}$. For $\mu\to 3e$, induced by vector operators, we get an suppression by a factor ${\cal O}(6)$, obtaining $m_*\gtrsim 0.5 \sqrt{g_*}\,{\rm TeV}$.

We have also studied the predictions of the model at numerical level. We have performed a random numerical scan over the ${\cal O}(1)$ parameters of the theory, namely the coefficients of the 5D Yukawa couplings, finding that the masses and mixing angles of the SM leptons can be naturally obtained. We have also studied the Wilson coefficients of CP and LFV operators, finding good agreement with the estimates.

The model provides a well motivated and computable framework for the study of an extended scalar sector. As we have shown, the radial components of $H$ and $\Phi$ mix in 5D, introducing corrections to the Higgs interactions. It would be interesting to study the phenomenology of the light scalars, as well the as the axion-like state in the present framework. Moreover, one can study the case of the Higgs as a hologram of $A_5$, in which case the Higgs potential is generated at loop level. We leave this subjects for future work.

Our analysis omitted higher-dimensional operators, besides those of ${\cal L}_y$. One could analyse the effect other higger dimensional operators, as for example corrections to fermionic kinetic terms. That study is beyond the scope of our work.

It is straightforward to extend the framework to include the quark sector.
An analysis in a two-site theory has been conducted in Ref.~\cite{DaRold:2021cca}.
It would be valuable to study the 5D description, as it offers greater predictive power.

%NLO corrections to dipole operators, see Neubert: dipoles from fermionic resonances.
\section*{Acknowledgments}
This work has been partially supported by CONICET, project PIP-11220200101426, and CNEA scholarships. We thank Esteban Roulet for discussions in the initial stages of the project.

\appendix
\section{KK decomposition of the scalar fields}\label{app-scalar}
In this appendix we describe the KK decomposition of the scalar fields.
\subsection{Vacuum expectation values}
The IR potential $V_1$ triggers spontaneous breaking of both the EW and FN symmetries, while preserving U(1)$_{\rm em}$, inducing the vevs $v_H(z)$ and $v_\Phi(z)$. Solving the bulk equations of motion for these vev profiles one gets:
\be
v_\chi=b_\chi z^{2-\beta_\chi} + c_\chi z^{2+\beta_\chi} \ , \qquad \beta_\chi=\sqrt{4+m_\chi^2L^2} \ ,  \qquad \chi=H,\Phi,
\label{eq-vevs}
\ee
where $b_\chi$ and $c_\chi$ are integration constants that depend on the BCs.

The UV and IR BCs are given by:
\be
\left.\left(\partial_zv_\chi+g_5^2a\frac{\partial}{\partial v_\chi} V_{i}\right)\right|_{z_{i}}=0 \ ,  \qquad \chi=H,\Phi, \qquad i=0,1.
\ee
where the UV BCs are evaluated at $z_i=z_0$ and the IR ones at $z_i=z_1$.

The UV BCs depend on the UV masses, we choose them to select the second solution of Eq.~(\ref{eq-vevs}):
\be
2+\beta_\chi+2g_5^2Lm_{0\chi}^2=0 \ ,  \qquad \chi=H,\Phi,
\ee 
leading to $b_\chi=0$.
%\be
%\left(\partial_zv_H+g_5^2a\frac{\partial}{\partial v_H} V_1\right)_{z_1}=0 \ ,
%\qquad
%\left(\partial_zv_\Phi+g_5^2a\frac{\partial}{\partial v_\Phi} V_1\right)_{z_1}=0
%\ee

The IR BCs are given by:
\ba
& \left[\partial_zv_H+g_5^2a(m_{1H}^2v_H+4\lambda_{1H}v_H^3+2\lambda_{1\Phi H}v_H^2v_\Phi)\right]_{z_1}=0 \ ,
\\
& \left[\partial_zv_\Phi+g_5^2a(m_{1\Phi}^2v_\Phi+4\lambda_{1\Phi}v_\Phi^3+2\lambda_{1\Phi H}v_\Phi^2v_H)\right]_{z_1}=0 \ .
\ea
For suitable choices of the IR masses and couplings $c_H$ and $c_\Phi$ do not vanish.

We find it useful to define an IR boundary mass: 
\be
\delta m_{1\chi}^2\equiv m_{1\chi}^2+\frac{2+\beta_\chi}{2g_5^2L} \ .
\label{eq-deltam1}
\ee
For $\delta m_{1\chi}^2=0$ and vanishing quartic couplings one obtains the usual conditions for the existence of a scalar 0-mode. 

Following Ref.~\cite{Cacciapaglia:2006mz} we define 4D vevs as the integrals of the 5D ones along the extra-dimension:
\be
v_{\chi 4}^2=\frac{1}{g_5^2}\int dz \ a(z)^3 v_\chi(z)^2 \ ,  \qquad \chi=H,\Phi.
\ee
Thus the 5D vevs are given by:
\be
v_\chi(z)=v_{\chi4}\frac{g_5}{\sqrt{L}}\frac{z_1}{L}\left(\frac{z}{z_1}\right)^{2+\beta_\chi} F(2+2\beta_\chi)\ ,
\ee
where the function $F(x)$ is defined by:
\be
F(x)=\left[\frac{x}{1-(z_0/z_1)^x}\right]^{1/2} \ .
\ee

The relation between the 4D vevs and the parameters of the IR potential is given by:
\ba
& v_{H4}=\frac{L}{z_1}\left(\frac{-2\lambda_{1\Phi}\delta m_{1H}^2+\lambda_{1\Phi H}\delta m_{1\Phi}^2}{4\lambda_{1\Phi}\lambda_{1H}-\lambda_{1\Phi H}^2}\right)^{1/2}\frac{1}{F(2+2\beta_H)} \ ,
\\
& v_{\Phi 4}=\frac{L}{z_1}\left(\frac{-2\lambda_{1H}\delta m_{1\Phi}^2+\lambda_{1\Phi H}\delta m_{1H}^2}{4\lambda_{1\Phi}\lambda_{1H}-\lambda_{1\Phi H}^2}\right)^{1/2}\frac{1}{F(2+2\beta_\phi)} \ .
\ea
The 4D vevs are expected to be of order $1/z_1$.

\subsection{Light states}
There are scalar physical states, that in general are massive and are a mixture of $H$ and $\Phi$. We consider in this section the lightest states, that would be 0-modes for $\delta m_{1\chi}^2=0$ and vanishing boundary quartic couplings. Below we present the KK decomposition of the scalar sector, as well as an approximation taking into account only the would be 0-modes, neglecting the mixing with the heavier KK states.

Expressing the 5D scalar fields as $H^t=(0,(v_H+h)/\sqrt{2})$ and $\Phi=(v_\Phi+\phi)/\sqrt{2}$, one obtains, for the physical fields $h$ and $\phi$, bulk equations of motion without mixing~\footnote{Tadpoles are also absent~\cite{Cacciapaglia:2006mz}.}:
\be
[(p^2+a^{-3}\partial_z a^3\partial_z-a^2m_\chi^2)\chi=0 \ ,  \qquad \chi=h,\phi.
\label{eq-5Ds}
\ee
The UV BCs are given by
\be
\left.\left[(\partial_z+2g_5^2am_{0\chi}^2)\chi\right]\right|_{z_0}=0 \ ,  \qquad \chi=h,\phi.
\label{eq-UVBCs}
\ee
On the other hand, the IR BCs mix both fields:
\ba
& \left.\{[\partial_z+a(2m_{1H}^2+12\lambda_{1H}v_H^2+2\lambda_{1\Phi H}v_\Phi^2)]h+4a\lambda_{1\Phi H}v_\Phi v_H \phi\}\right|_{z_1} \ ,
\label{eq-IRBCs1}
\\
& \left.\{[\partial_z+a(2m_{1\Phi}^2+12\lambda_{1\Phi}v_\Phi^2+2\lambda_{1\Phi H}v_H^2)]\phi+4a\lambda_{1\Phi H}v_\Phi v_H h\}\right|_{z_1} \ .
\label{eq-IRBCs2}
\ea
%
%\ba
%& \left.[\partial_z h+a(2m_{1H}^2h+12\lambda_{1H}v_H^2h+2\lambda_{1\Phi H}v_\Phi^2h+4\lambda_{1\Phi H}v_\Phi v_H \phi)]\right|_{z_1} \ ,
%\\
%& \left.[\partial_z \phi+a(2m_{1\Phi}^2\phi+12\lambda_{1\Phi}v_\Phi^2\phi+2\lambda_{1\Phi H}v_H^2\phi+4\lambda_{1\Phi H}v_\Phi v_H h)]\right|_{z_1} \ .
%\ea
We decompose the 5D scalars fields in KK modes as:
\be
h=\sum_n s^{(n)}(x^\mu)f^h_n(z) \ , \qquad \phi=\sum_n s^{(n)}(x^\mu)f^\phi_n(z) \ , 
\label{eq-KKs}
\ee
with the normalization condition:
\be
\frac{1}{g_5^2}\int dz\ a(z)^3[f^h_n(z)f^h_m(z)+f^\phi_n(z)f^\phi_m(z)]=\delta_{nm}
\ee
and KK wave functions given by:
\be
f^h_n(z)=\frac{z^2}{N^h_n}[J_{\beta_H}(m_n z)+b^h_n Y_{\beta_H}(m_n z)] 
\ , \qquad 
f^\phi_n(z)=\frac{z^2}{N^\phi_n}[J_{\beta_\phi}(m_n z)+b^\phi_n Y_{\beta_\phi}(m_n z)] \ ,
\ee
where $b_n^{h,\phi}$ are integration constants, $N_n^{h,\phi}$ are normalization factors and we have used $p^2=m_n^2$ in Eq.~(\ref{eq-5Ds}). Since $V_1$ induces masses for the KK modes, in general the sums of Eqs.~(\ref{eq-KKs}) run over $n>0$.

One can use, for example, Eqs.~(\ref{eq-UVBCs}) to obtain the integration constants. The spectrum of scalar states, $m_n$, can then be obtained from Eqs.~(\ref{eq-IRBCs1}) and~(\ref{eq-IRBCs2}), and can be written as:
\be
\left.(\partial_z f^h_n+\tilde m_{1h} f^h_n)(\partial_z f^\phi_n+\tilde m_{1\phi} f^\phi_n)\right|_{z_1}=\left.\tilde m_{1\phi h}\tilde m_{1h\phi} f^h_n f^\phi_n\right|_{z_1} \ ,
\label{eq-specs}
\ee
where:
\ba
& \tilde m_{1h} = \left.a(2m_{1H}^2+12\lambda_{1H}v_H^2+2\lambda_{1\Phi H}v_\Phi^2)\right|_{z_1} \ ,
\\
& \tilde m_{1\phi} = \left.a(2m_{1\Phi}^2+12\lambda_{1\Phi}v_\Phi^2+2\lambda_{1\Phi H}v_H^2)\right|_{z_1} \ ,
\\
& \tilde m_{1\phi h} = \tilde m_{1h\phi} = \left.4a \lambda_{1\Phi H}v_\Phi v_H h\right|_{z_1}  \ .
\ea
Notice that Eq.~(\ref{eq-specs}) does not depend on the normalization factor of the KK wave functions. For $\lambda_{1\Phi H}=0$ the mixing vanishes and Eq.~(\ref{eq-specs}) has two sets of solutions that reproduce the spectrum of two independent scalar fields.

One can compute the spectrum of scalar states by considering a perturbative expansion in insertions of the IR boundary terms, the so called zero mode approximation. We first solve for the 0-modes obtained when $\delta m_{1H}^2=\delta m_{1\Phi}^2=0$ and all the quartics vanish: $\lambda_{1H}=\lambda_{1\Phi}=\lambda_{1\Phi H}=0$, and then we define a mass matrix of the 0-modes that accounts for the effect of those terms of the IR potential. 
The 0-modes are given by:
\be
f^\chi_0(z)=\frac{g_5}{\sqrt{L}}\frac{z_1}{L}\left(\frac{z}{z_1}\right)^{2+\beta_\chi} F(2+2\beta_\chi)\ ,  \qquad \chi=h,\phi.
\ee

\begin{figure}[t]
\centering
\includegraphics[width=0.49\textwidth]{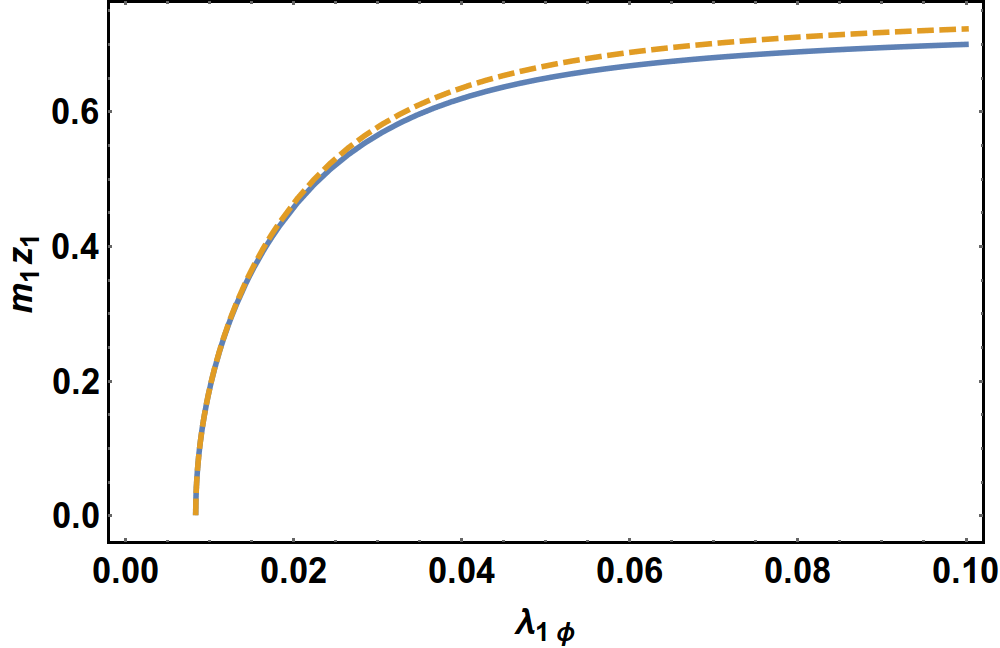}
\includegraphics[width=0.49\textwidth]{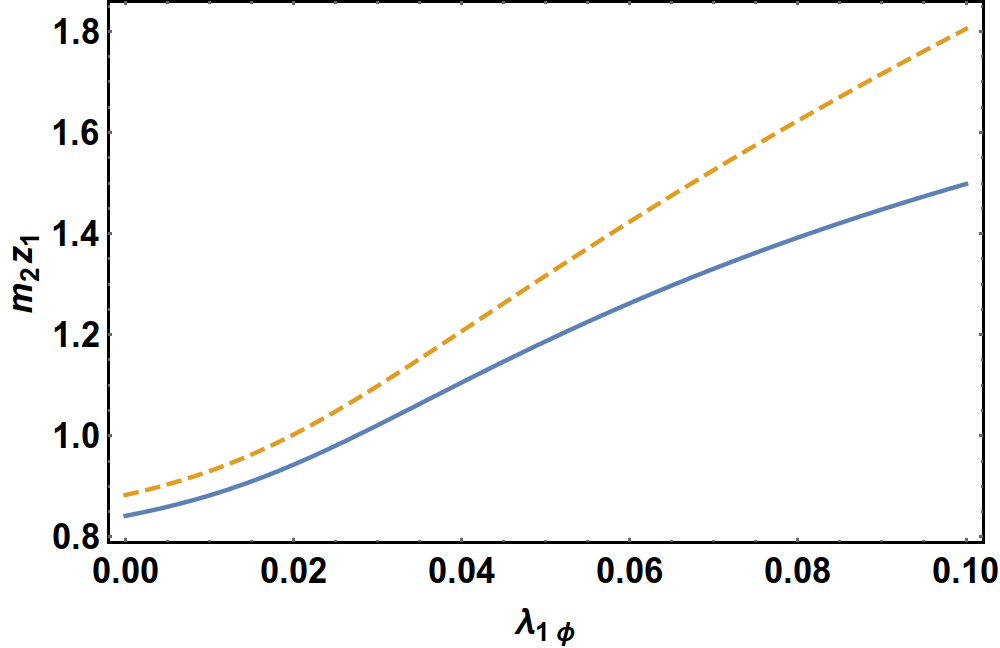}
\caption{Masses of the first (left panel) and second (right panel) KK scalar modes as function of the quartic coupling $\lambda_{1\Phi}$ for the BM point, in solid blue we show the full expression of Eq.~(\ref{eq-specs}) and in dashed orange the formula derived from the approximation of Eq.~(\ref{eq-M2}).}
\label{fig-KKms}
\end{figure}

In terms of these 0-modes one can define a mass matrix as follows: 
\be
M_s^2=2 a^4\left.\left(
\begin{array}{cc} 
(\delta m_{1H}^2+6\lambda_{1H}v_H^2+\lambda_{1\Phi H}v_\Phi^2)(f^h_0)^2 & 2\lambda_{1\Phi H}v_\Phi v_H f^h_0 f^\phi_0
\\
2\lambda_{1\Phi H}v_\Phi v_H f^h_0 f^\phi_0 & (\delta m_{1\Phi}^2+6\lambda_{1\Phi}v_\Phi^2+\lambda_{1\Phi H}v_H^2)(f^\phi_0)^2
\end{array}
\right)\right|_{z_1} \ .
\label{eq-M2}
\ee
This approximation does not take into account the mixing with the massive modes induced by the IR potential, however for the region of the parameter space we are interested in the corrections are small, as can be seen in Fig.~\ref{fig-KKms}, for a suitable choice of parameters. For $m_1$ the differences between the full calculation and the approximation are below $3\%$, whereas for $m_2$ they are below $20\%$.
As usual, obtaining a suppression in the Higgs mass requires tuning of the quartic coupling.% More insight on this problem can be obtained by considering the approximations discussed below.

\section{Fermion matrix diagonalization}
\label{ap-diag}
We consider first the simple case of just one generation. From Eq.~(\ref{eq-ye}) the mass matrix of the charged leptons is~(\ref{eq-me1}):
\ba
M_{E}&=\left(\begin{array}{ccc} v_{H4}y_{11} & 0 & v_{H4}y_{13} \\ v_{H4}y_{21} & m_1^{\rm L} & v_{H4}y_{23} \\ 0 & v_{H4}y_{32} & m_1^E \end{array}\right) \ ,\nonumber 
\\
&\sim  \left(\begin{array}{ccc} \frac{g_5}{\sqrt{L}}v_{H4}F(1-2c_{\rm L})F(1+2c_E)\delta^{\alpha_E} & 0 & \frac{g_5}{\sqrt{L}}v_{H4}F(1-2c_{\rm L})\delta^{\alpha_E} \\ \frac{g_5}{\sqrt{L}}v_{H4}F(1+2c_E)\delta^{\alpha_E} & m_1^{\rm L} & \frac{g_5}{\sqrt{L}}v_{H4}\delta^{\alpha_E} \\ 0 & \frac{g_5}{\sqrt{L}}v_{H4}\delta^{\alpha_E} & m_1^E \end{array}\right)\ ,
\label{eq-me2}
\ea
where $m_1^{\rm L}$ and $m_1^{E}$ are the masses of the first KK states, respectively ${\rm L}^{(1)}$ and $E^{(1)}$. Since $\delta^{\alpha_E}\ll 1$ only if $\alpha_E>0$, a condition that is not always satisfied once flavor is introduced, we will not expand in powers of $\delta$. Instead we will expand in powers of $F(1-2c_{\rm L})$ and $F(1+2c_E)$.

We consider $M_E$ to zeroth order in $F(1\mp 2c_{{\rm L},E})$
\be
M_E^{(0)}=\left(\begin{array}{ccc} 0 & 0 & 0 \\ 0 & m_1^{\rm L} & v_{H4}y_{23} \\ 0 & v_{H4}y_{32} & m_1^E \end{array}\right) \ , 
\ee
and we perform a diagonalization of $M_E^{(0)}$ with a bi-unitary transformation:
\be
M_E^{(1)}= U_L^{(0)\dagger} M_E^{(0)} U_R^{(0)} \ ,
\ee
where
\be 
M_E^{(1)}={\rm diag}(0,m_1',m_1'') \ ,
\qquad
U_{L,R}^{(0)}=\left(\begin{array}{ccc} 1 & 0 & 0 \\ 0 & \cos\theta_{L,R}^{(0)} & -\sin\theta_{L,R}^{(0)} \\ 0 & \sin\theta_{L,R}^{(0)} & \cos\theta_{L,R}^{(0)} \end{array}\right) \ .
\ee
It is straightforward to compute $m_1'$, $m_1''$ and $\sin\theta_{L,R}^{(0)}$. 

Armed with these unitary matrices we perform a rotation of $M_E$ as follows:
\be
M_E'= U_L^{(0)\dagger} M_E U_R^{(0)}=\left(\begin{array}{ccc} v_{H4}y_{11} & \sin\theta_R^{(0)}v_{H4}y_{13} & \cos\theta_R^{(0)}v_{H4}y_{13} \\ \cos\theta_L^{(0)}v_{H4}y_{21} & m_1' & 0 \\ -\sin\theta_L^{(0)}v_{H4}y_{21} & 0 & m_1'' \end{array}\right)  \ ,
\ee
and we diagonalize $M_E'$ with a new bi-unitary transformation, calling $\tilde M_E$ the diagonal mass matrix:
\be
\tilde M_E= U_L'^{\dagger} M_E' U_R'\ .
\ee

The matrix of Yukawa couplings is given by:
\be
\tilde y_E= U_L'^\dagger U_L^{(0)\dagger} y_E U_R^{(0)}U_R' \ .
\ee
with $y_E$ given in Eq.~(\ref{eq-ye}), restricted to one generation. The matrices of couplings with the first KK vector are:
\be
\tilde g_L= U_L'^\dagger U_L^{(0)\dagger} g_L U_L^{(0)}U_L' \ ,
\qquad
\tilde g_R= U_R'^\dagger U_R^{(0)\dagger} g_R U_R^{(0)}U_R' \ .
\ee

It is straightforward to obtain the eigenvalues, the rotation matrices and the matrices of couplings performing an expansion in powers of $F(1\mp 2c_{{\rm L},E})$. Simple expressions can be obtained expanding also in powers of $v_{H4}$. Taking the limit $m_1^{\rm L}=m_1^E\equiv m_{\rm KK}$, the eigenvalues are given by:
\ba
&\tilde m_1\approx v_{H4}y_{11} +\frac{v_{H4}^3}{m_{\rm KK}^2}y_{13}y_{32}y_{21} +\dots\ , \nonumber
\\
&\tilde m_{2,3}\approx m_{\rm KK}\mp\frac{1}{2}v_{H4}(y_{23}+y_{32})+\frac{v_{H4}^2}{8m_{\rm KK}}(y_{23}-y_{32})^2+\frac{v_{H4}^2}{4m_{\rm KK}}(y_{13}^2+y_{21}^2)+\dots
\label{eq-lam}
\ea
where $m_1$ is the mass of the would be 0-modes, and $m_{2,3}$ are the masses of the first level of resonances. We have expanded to second order in $v_{H4}/m_{\rm KK}$ and we show also the leading order correction in $F(1\mp 2c_{{\rm L},E})$ for the resonances.
The rotation matrices can be expanded as:
\ba
&U_L'\approx\left(
\begin{array}{ccc}
 1-\frac{v_{H4}^2 y_{13}^2}{2 m_{\text{KK}}^2} & \frac{ v_{H4}y_{13}}{\sqrt{2} m_{\text{KK}}}+\frac{y_{13} \left(y_{23}+3 y_{32}\right) v_{H4}^2}{4 \sqrt{2} m_{\text{KK}}^2} &
   \frac{v_{H4}y_{13}}{\sqrt{2} m_{\text{KK}}}-\frac{v_{H4}^2y_{13} \left(y_{23}+3 y_{32}\right)}{4 \sqrt{2} m_{\text{KK}}^2} \\
 -\frac{v_{H4}y_{13}}{\sqrt{2} m_{\text{KK}}}-\frac{v_{H4}^2y_{13} \left(y_{23}+3 y_{32}\right)}{4 \sqrt{2} m_{\text{KK}}^2} & 1-\frac{v_{H4}^2 y_{13}^2}{4 m_{\text{KK}}^2} &
   \frac{v_{H4} \left(y_{13}^2-y_{21}^2\right)}{4 m_{\text{KK}} y_{23}+4 m_{\text{KK}} y_{32}}-\frac{v_{H4}^2 y_{13}^2}{4 m_{\text{KK}}^2} \\
 -\frac{v_{H4}y_{13}}{\sqrt{2} m_{\text{KK}}}+\frac{v_{H4}^2 y_{13} \left(y_{23}+3 y_{32}\right)}{4 \sqrt{2} m_{\text{KK}}^2} & \frac{v_{H4} \left(y_{21}^2-y_{13}^2\right)}{4
   m_{\text{KK}} \left(y_{23}+y_{32}\right)}-\frac{v_{H4}^2 y_{13}^2}{4 m_{\text{KK}}^2} & 1-\frac{v_{H4}^2 y_{13}^2}{4 m_{\text{KK}}^2} \\
\end{array}
\right) \nonumber \\
&U'_R\approx\left(
\begin{array}{ccc}
 1-\frac{v_{H4}^2 y_{21}^2}{2 m_{\text{KK}}^2} & -\frac{v_{H4}y_{21}}{\sqrt{2} m_{\text{KK}}} -\frac{v_{H4}^2y_{21} \left(y_{23}+3 y_{32}\right)}{4 \sqrt{2} m_{\text{KK}}^2} &
   \frac{v_{H4} y_{21}}{\sqrt{2} m_{\text{KK}}}-\frac{v_{H4}^2 y_{21} \left(y_{23}+3 y_{32}\right)}{4 \sqrt{2} m_{\text{KK}}^2} \\
 \frac{v_{H4}y_{21}}{\sqrt{2} m_{\text{KK}}}+\frac{v_{H4}^2y_{21} \left(y_{23}+3 y_{32}\right)}{4 \sqrt{2} m_{\text{KK}}^2} & 1-\frac{v_{H4}^2 y_{21}^2}{4 m_{\text{KK}}^2} &
  \frac{v_{H4} \left(y_{13}^2-y_{21}^2\right)}{4 m_{\text{KK}} y_{23}+4 m_{\text{KK}} y_{32}}+ \frac{v_{H4}^2 y_{21}^2}{4 m_{\text{KK}}^2} \\
 -\frac{v_{H4} y_{21}}{\sqrt{2} m_{\text{KK}}}+\frac{v_{H4}^2 y_{21} \left(y_{23}+3 y_{32}\right)}{4 \sqrt{2} m_{\text{KK}}^2} & \frac{v_{H4} \left(y_{21}^2-y_{13}^2\right)}{4 m_{\text{KK}} \left(y_{23}+y_{32}\right)} + \frac{v_{H4}^2 y_{21}^2}{4 m_{\text{KK}}^2} & 1-\frac{v_{H4}^2 y_{21}^2}{4 m_{\text{KK}}^2} \\
\end{array}
\right) \ .
\ea

Expanding the couplings to first order in $v_{H4}/m_{\rm KK}$ and to second order in $F(1\mp 2c_{{\rm L},E})$ we get:
\ba
& \tilde y\approx
\left(
\begin{array}{ccc}
y_{11} 
& \scriptstyle \frac{y_{13}}{\sqrt{2}}-\frac{v_{H4} y_{13} \left(y_{23}-5 y_{32}\right)}{4 \sqrt{2} m_{\text{KK}}} 
& \scriptstyle \frac{v_{H4} \left(y_{23}-5 y_{32}\right) y_{13}}{4 \sqrt{2} m_{\text{KK}}}+\frac{y_{13}}{\sqrt{2}} 
\\
\scriptstyle -\frac{y_{21}}{\sqrt{2}} + \frac{v_{H4} y_{21} \left(y_{23}-5 y_{32}\right)}{4 \sqrt{2} m_{\text{KK}}}
& \scriptstyle -\frac{y_{23}+y_{32}}{2} +v_{H4} \left(\frac{\left(y_{23}-y_{32}\right){}^2}{4 m_{\text{KK}}}+\frac{y_{13}^2+y_{21}^2}{2 m_{\text{KK}}}\right) 
& \scriptstyle \frac{y_{32}-y_{23}}{2}  + \frac{v_{H4} \left(y_{13}-y_{21}\right) \left(y_{13}+y_{21}\right)}{4 m_{\text{KK}}} 
\\
\scriptstyle \frac{y_{21}}{\sqrt{2}} + \frac{v_{H4} \left(y_{23}-5 y_{32}\right) y_{21}}{4 \sqrt{2} m_{\text{KK}}} 
& \scriptstyle \frac{y_{23}-y_{32}}{2}  + \frac{v_{H4} \left(y_{13}^2-y_{21}^2\right)}{4 m_{\text{KK}}} 
& \scriptstyle \frac{y_{23}+y_{32}}{2} + v_{H4} \left(\frac{\left(y_{23}-y_{32}\right){}^2}{4 m_{\text{KK}}}+\frac{y_{13}^2+y_{21}^2}{2 m_{\text{KK}}}\right)
\end{array}
\right) \ ,
\nonumber \\
& \tilde g_L\approx
\left(
\begin{array}{ccc}
 g_{L,11} 
& \dots %\scriptstyle v_{\text{H4}} \left(\frac{\left(y_{23}-y_{32}\right) g_{L,12}}{4 \sqrt{2} m_{\text{KK}}}+\frac{\left(y_{21}^2-y_{13}^2\right) g_{L,12}}{4 \sqrt{2} m_{\text{KK}} \left(y_{23}+y_{32}\right)}+\frac{y_{13} \left(g_{L,11}-g_{L,33}\right)}{\sqrt{2} m_{\text{KK}}}\right)-\frac{g_{L,12}}{\sqrt{2}} 
& \dots %\scriptstyle \frac{g_{L,12}}{\sqrt{2}}+v_{\text{H4}} \left(\frac{\left(y_{23}-y_{32}\right) g_{L,12}}{4 \sqrt{2} m_{\text{KK}}}+\frac{\left(y_{21}^2-y_{13}^2\right) g_{L,12}}{4 \sqrt{2} m_{\text{KK}} \left(y_{23}+y_{32}\right)}+\frac{y_{13} \left(g_{L,11}-g_{L,33}\right)}{\sqrt{2} m_{\text{KK}}}\right) 
\\
\scriptstyle -\frac{g_{L,12}}{\sqrt{2}} + v_{\text{H4}} \left(\frac{\left(y_{23}-y_{32}\right) g_{L,12}}{4 \sqrt{2} m_{\text{KK}}}+\frac{\left(y_{21}^2-y_{13}^2\right) g_{L,12}}{4 \sqrt{2} m_{\text{KK}}
   \left(y_{23}+y_{32}\right)}+\frac{y_{13} \left(g_{L,11}-g_{L,33}\right)}{\sqrt{2} m_{\text{KK}}}\right)
& \dots %\scriptstyle \frac{1}{2} \left(g_{L,22}+g_{L,33}\right)+v_{\text{H4}} \left(-\frac{y_{13} g_{L,12}}{m_{\text{KK}}}+\frac{\left(y_{23}-y_{32}\right) \left(g_{L,33}-g_{L,22}\right)}{4 m_{\text{KK}}}-\frac{\left(y_{13}-y_{21}\right) \left(y_{13}+y_{21}\right) \left(g_{L,33}-g_{L,22}\right)}{4 m_{\text{KK}} \left(y_{23}+y_{32}\right)}\right) 
& \dots %\scriptstyle \frac{1}{2} \left(g_{L,33}-g_{L,22}\right) 
\\
\scriptstyle \frac{g_{L,12}}{\sqrt{2}}+v_{\text{H4}} \left(\frac{\left(y_{23}-y_{32}\right) g_{L,12}}{4 \sqrt{2} m_{\text{KK}}}+\frac{\left(y_{21}^2-y_{13}^2\right) g_{L,12}}{4 \sqrt{2} m_{\text{KK}}
   \left(y_{23}+y_{32}\right)}+\frac{y_{13} \left(g_{L,11}-g_{L,33}\right)}{\sqrt{2} m_{\text{KK}}}\right) 
& \dots %\scriptstyle \frac{1}{2} \left(g_{L,33}-g_{L,22}\right) 
& \dots %\scriptstyle \frac{1}{2}\left(g_{L,22}+g_{L,33}\right)+v_{\text{H4}} \left(\frac{y_{13} g_{L,12}}{m_{\text{KK}}}-\frac{\left(y_{23}-y_{32}\right) \left(g_{L,33}-g_{L,22}\right)}{4 m_{\text{KK}}}+\frac{\left(y_{13}-y_{21}\right) \left(y_{13}+y_{21}\right) \left(g_{L,33}-g_{L,22}\right)}{4 m_{\text{KK}} \left(y_{23}+y_{32}\right)}\right)
\end{array}
\right)
\ ,
\nonumber \\
& \tilde g_R\approx
\left(
\begin{array}{ccc}
 g_{R,11} 
& \dots %\frac{g_{R,13}}{\sqrt{2}}+v_{\text{H4}} \left(\frac{\left(y_{32}-y_{23}\right) g_{R,13}}{4 \sqrt{2} m_{\text{KK}}}+\frac{\left(y_{21}^2-y_{13}^2\right) g_{R,13}}{4 \sqrt{2} m_{\text{KK}} \left(y_{23}+y_{32}\right)}+\frac{y_{21} \left(g_{R,22}-g_{R,11}\right)}{\sqrt{2} m_{\text{KK}}}\right) 
& \dots %\frac{g_{R,13}}{\sqrt{2}}+v_{\text{H4}} \left(\frac{\left(y_{23}-y_{32}\right) g_{R,13}}{4 \sqrt{2} m_{\text{KK}}}+\frac{\left(y_{13}-y_{21}\right) \left(y_{13}+y_{21}\right) g_{R,13}}{4 \sqrt{2} m_{\text{KK}} \left(y_{23}+y_{32}\right)}+\frac{y_{21} \left(g_{R,11}-g_{R,22}\right)}{\sqrt{2} m_{\text{KK}}}\right) 
\\
\scriptstyle  \frac{g_{R,13}}{\sqrt{2}}+v_{\text{H4}} \left(\frac{\left(y_{32}-y_{23}\right) g_{R,13}}{4 \sqrt{2} m_{\text{KK}}}+\frac{\left(y_{21}^2-y_{13}^2\right) g_{R,13}}{4 \sqrt{2} m_{\text{KK}} \left(y_{23}+y_{32}\right)}+\frac{y_{21} \left(g_{R,22}-g_{R,11}\right)}{\sqrt{2} m_{\text{KK}}}\right) 
& \dots %\scriptstyle \frac{1}{2} \left(g_{R,22}+g_{R,33}\right)+v_{\text{H4}} \left(-\frac{y_{21} g_{R,13}}{m_{\text{KK}}}-\frac{\left(y_{23}-y_{32}\right) \left(g_{R,33}-g_{R,22}\right)}{4 m_{\text{KK}}}-\frac{\left(y_{13}-y_{21}\right) \left(y_{13}+y_{21}\right) \left(g_{R,33}-g_{R,22}\right)}{4 m_{\text{KK}} \left(y_{23}+y_{32}\right)}\right) 
& \dots %\frac{1}{2} \left(g_{R,33}-g_{R,22}\right) 
\\ \scriptstyle \frac{g_{R,13}}{\sqrt{2}}+v_{\text{H4}} \left(\frac{\left(y_{23}-y_{32}\right) g_{R,13}}{4 \sqrt{2} m_{\text{KK}}}+\frac{\left(y_{13}-y_{21}\right) \left(y_{13}+y_{21}\right) g_{R,13}}{4 \sqrt{2} m_{\text{KK}} \left(y_{23}+y_{32}\right)}+\frac{y_{21} \left(g_{R,11}-g_{R,22}\right)}{\sqrt{2} m_{\text{KK}}}\right) 
& \dots %\frac{1}{2} \left(g_{R,33}-g_{R,22}\right) 
& \dots %\frac{1}{2} \left(g_{R,22}+g_{R,33}\right)+v_{\text{H4}} \left(\frac{y_{21} g_{R,13}}{m_{\text{KK}}}+\frac{\left(y_{23}-y_{32}\right) \left(g_{R,33}-g_{R,22}\right)}{4 m_{\text{KK}}}+\frac{\left(y_{13}-y_{21}\right) \left(y_{13}+y_{21}\right) \left(g_{R,33}-g_{R,22}\right)}{4 m_{\text{KK}} \left(y_{23}+y_{32}\right)}\right) 
\end{array}
\right) \ ,
\ea
where for $\tilde g_{L,R}$, that are symmetric, we show only the first column, that involves the lightest state. The couplings between KK states are not shown since they are not needed in this article and they involve rather long expressions.

Using these results it is straightforward to compute the coefficients of dipole~(\ref{eq-Wdip}) and vector~(\ref{eq-Wvec}) operators, obtaining Eq.~(\ref{eq-WCs}).

When flavor is taken into account, the couplings and the KK masses become $3\times 3$ matrices, with $g_{L,jk}$ and $g_{R,jk}$, as well as $m_{\rm KK}$ being diagonal, whereas the Yukawa couplings $y_{jk}$ have a non-trivial structure, Eq.~(\ref{eq-yejk}). As an example, the second term of $\tilde m_1$ in Eq.~(\ref{eq-lam}) can be written as: 
\be
[v_{H4}^3 y_{13}(m^E_1)^{-1}y_{32}(m^{\rm L}_1)^{-1}y_{21}]^{jk} \ ,
\ee
with $j,k$ being generation indices.

Finally, to go to the mass basis one has to perform a bi-unitary transformation in generation space. As described in sec.~\ref{sec-lightleptons}, we consider the limit where the masses of the would be 0-modes are approximated by the leading order term in powers of $v_{H4}/m_{\rm KK}$ and $F(1\mp 2c)$, as in Eq.~(\ref{eq-m0modes}). 

Below we consider the diagonalization in flavor space of the mass matrix of the charged leptons of the SM for BM1. We express the factors $\delta^{\alpha_E^{jk}}$ and $F(1\mp 2c_{{\rm L},E}^j)$ in terms of powers of $\lambda_C$ using Eqs.~(\ref{eq-deltan}) and (\ref{eq-Fn}):
\be
m_E\approx \frac{g_{5}}{\sqrt{L}}v_{H4} \left(\begin{array}{ccc} x_E^{11}\lambda_C^9 & x_E^{12}\lambda_C^{10} & x_E^{13}\lambda_C^{12} \\ x_E^{21}\lambda_C^9 & x_E^{22} \lambda_C^6 & x_E^{23}\lambda_C^{8} \\ x_E^{31}\lambda_C^9 & x_E^{32}\lambda_C^6 & x_E^{33}\lambda_C^4 \end{array}\right) \ .
\label{eq-Ye}
\ee

The eigenvalues, that reproduce the masses of the charged leptons of the SM, are given to leading order by:
\be
m_i=\frac{g_{5}}{\sqrt{L}}v_{H4} x_E^{ii}\lambda_C^{\omega_i} \ , \qquad \omega_i =9,6,4, \qquad {\rm for} \ i=1,2,3,
\ee
and the Left- and Right-handed rotation matrices are:
\ba
& U_{e_L}\approx \left(\begin{array}{ccc} 
1 & \frac{x_E^{12}}{x_E^{22}}\lambda_C^{4} & \frac{x_E^{12}x_E^{32}+x_E^{13}x_E^{33}}{(x_E^{33})^2}\lambda_C^{8}
\\ -\frac{x_E^{12}}{x_E^{22}}\lambda_C^{4} & 1 & \frac{x_E^{22}x_E^{32}+x_E^{23}x_E^{33}}{(x_E^{33})^2}\lambda_C^{4} 
\\ \frac{-x_E^{13}x_E^{22}+x_E^{12}x_E^{23}}{x_E^{22}x_E^{33}}\lambda_C^{8} & -\frac{x_E^{22}x_E^{32}+x_E^{23}x_E^{33}}{(x_E^{33})^2}\lambda_C^{4} & 1 
\end{array}\right) \ ,
\nonumber \\
& U_{e_R}\approx \left(\begin{array}{ccc} 
1 & \frac{x_E^{21}}{x_E^{22}}\lambda_C^{3} & \frac{x_E^{31}}{x_E^{33}}\lambda_C^{5}
\\ -\frac{x_E^{21}}{x_E^{22}}\lambda_C^{3} & 1 & \frac{x_E^{22}}{x_E^{33}}\lambda_C^{2} 
\\ \frac{-x_E^{31}x_E^{22}+x_E^{21}x_E^{32}}{x_E^{22}x_E^{33}}\lambda_C^{5} & -\frac{x_E^{32}}{x_E^{33}}\lambda_C^{2} & 1 
\end{array}\right) \ ,
\label{eq-ueap}
\ea
where to simplify the expressions we have assumed real coefficients. 

Armed with these matrices one can obtain the Wilson coefficients of sec.~\ref{sec-LFV}. In our calculations of the Wilson coefficients we have used expressions that contain higher orders in the expansion in powers of $\lambda_C$.

%%%%%%%%%%%%%%%%%%%%%%%%%%%%%%%%%%%%%%%%%%%%%%%%%%%%%%%%%%%%%%%%

\bibliographystyle{JHEP}
\bibliography{biblio}

\end{document}